\documentclass[submission, Phys]{SciPost}
\usepackage[english]{babel}
\usepackage[utf8]{inputenc}
\usepackage{amsfonts}
\usepackage[T1]{fontenc}
\usepackage[titletoc,title]{appendix}
\usepackage{epstopdf}
\usepackage{amsmath} 
\usepackage{amssymb}
\usepackage{dsfont}
\usepackage{braket}
\usepackage{mathtools}
\usepackage{xcolor}
\usepackage{tikz}
\usepackage[export]{adjustbox}
\usepackage{hyperref}
\usepackage{bbold}
\usepackage[normalem]{ulem}
\usepackage{adjustbox}
\hypersetup{
    unicode=false, 
    pdftoolbar=false, 
    pdfmenubar=true, 
    pdffitwindow=false, 
    pdfstartview={}, 
    pdftitle={Periodic orbits and quantum many-body scars in integrable spin chains}, 
    pdfauthor={E. Petrova et al.}, 
    pdfsubject={}, 
    pdfcreator={}, 
    pdfproducer={}, 
    pdfkeywords={MPS, tangent space, time evolution, periodic orbits, integrability, Bethe ansatz}, 
    pdfnewwindow=true, 
    colorlinks=true, 
    linkcolor=black, 
    citecolor=black, 
    filecolor=black, 
    urlcolor=black 
}
\usepackage{amsmath}
\usepackage{algpseudocodex}
\usepackage{newfloat}
\DeclareFloatingEnvironment[name=Algorithm]{algorithm}

\usepackage{outline}

\newcommand{\im}{{\text{Im}}}

\makeatletter
\newcommand{\ostar}{\mathbin{\mathpalette\make@circled\star}}
\newcommand{\make@circled}[2]{
  \ooalign{$\m@th#1\smallbigcirc{#1}$\cr\hidewidth$\m@th#1#2$\hidewidth\cr}
}
\newcommand{\smallbigcirc}[1]{
  \vcenter{\hbox{\scalebox{0.77778}{$\m@th#1\bigcirc$}}}
}
\makeatother

\begin{document}
\title{Periodic orbits and quantum many-body scars in integrable spin chains}

\author{
\begin{minipage}{0.95\textwidth}
\centering
Elena Petrova$^{1}$,
Maksym Serbyn$^{1}$,
Marko Ljubotina$^{2,3}$\\[1.5mm]
{\footnotesize
$^{1}$ Institute of Science and Technology Austria (ISTA), Am Campus 1, 3400 Klosterneuburg, Austria\\
$^{2}$ Physics Department, Technical University of Munich, TUM School of Natural Sciences, Lichtenbergstr. 4, 85748 Garching, Germany\\
$^{3}$ Munich Center for Quantum Science and Technology (MCQST), Schellingstr. 4, 80799 M\"unchen, Germany
}
\end{minipage}
}
\maketitle

    \begin{abstract}
        Quantum many-body scars provide an exception to generic thermalising dynamics, but their relation to integrability remains unclear. Here we apply the periodic-orbit framework to the integrable XYZ/XXZ spin-1/2 chain, developing an energy-resolved approach that enables us to locate and track periodic orbits across the spectrum. We identify families of scarred orbits and resolve their supporting eigenstates in terms of Bethe-ansatz quantum numbers, revealing a common string core dressed by zero-momentum magnon pairs. This allows us to reconstruct both the scarred eigenstates and the associated towers directly within the Bethe ansatz, and explain the equidistant tower spacing analytically from the decoupling of zero-momentum magnon pairs in the Bethe equations, providing a microscopic realisation of scar phenomenology in an integrable setting. We further show that this structure persists upon breaking integrability with a transverse field, with the periodic orbits continuing to govern the dynamics. Our results establish a direct connection between the periodic-orbit picture of quantum scarring and the algebraic structure of integrable models, showing that key features of scar dynamics can be understood analytically within the Bethe ansatz framework.
        
    \end{abstract}

    \section{Introduction}    

    Generic many-body Hamiltonian dynamics drives entanglement growth and thermalisation: a typical initial state quickly develops volume-law entanglement, local observables relax to thermal values, and all memory of the initial condition is lost~\cite{deutsch1991quantum, srednicki1994chaos, rigol2008thermalization}. In sharp contrast to this generic behaviour, quantum many-body scars~\cite{Turner2017, shiraishi2017systematic, Serbyn:2021vc, Moudgalya_2022, chandran23} stand out as a striking exception. First identified numerically in the kinetically constrained PXP model~\cite{Turner2017} and observed experimentally as long-lived coherent oscillations in Rydberg-atom arrays~\cite{bernien2017probing}, special states, called scarred initial states, revive periodically under time evolution rather than relaxing and maintain low entanglement entropy. This behaviour traces back to a structural feature of the spectrum: the weight of a scarred initial state is concentrated on a sub-extensive set of approximately equidistant (scarred) eigenstates, such that the dynamics is effectively confined to a low-dimensional subspace, with revivals following from the near-equidistant spacing of the scarred eigenstates. In the PXP model, the supporting eigenstates themselves exhibit anomalously low entanglement relative to their thermal neighbours~\cite{Turner2017}, though this is a model-specific observation rather than part of the general picture~\cite{Serbyn:2021vc, Moudgalya_2022, chandran23}. A large body of subsequent works has established scar phenomenology across a wide range of models, including systems with exact~\cite{shiraishi2017systematic} and approximately~\cite{Turner2017, LjubotinaPRXQ} equidistant towers (sets of states with approximately the same energy and high overlaps with the N\'eel states), algebraic constructions~\cite{Moudgalya_2022}, and fragmented Hilbert spaces~\cite{chandran23}.
    
    Beyond spectral diagnostics, quantum many-body scars admit a clear dynamical characterisation that has been developed along two complementary lines. In the semiclassical picture~\cite{HummelPRL, Dag24PRL, ermakov2024, pizzi2024, Dag24}, scars appear as quantum signatures of unstable periodic orbits of an underlying mean-field or projected dynamics, in direct analogy with single-particle quantum scars~\cite{heller1984bound}. In the variational formulation~\cite{ho2019periodic, michailidis2020slow, LjubotinaPRXQ}, which provides a generalisation of the semiclassical approach, closed periodic orbits of the time-dependent variational principle (TDVP~\cite{dirac1930note, lubich2008quantum, haegeman2011time}) on a low bond dimension matrix product state (MPS) manifold~\cite{vidal2003efficient,vidal2004efficient, vidal2007classical} give rise directly to long-lived coherent dynamics associated with scars. The leakage of these projected orbits outside the manifold measures how faithfully the projected dynamics approximates the true Schr\"odinger evolution, and stays small precisely when the supporting eigenstates carry low enough entanglement to fit within the chosen bond dimension $\chi$ MPS manifold. Both viewpoints place periodic orbits within an effectively low-dimensional manifold at the heart of the phenomenon, and together they suggest a constructive route to finding scars: search directly for such orbits on an MPS manifold rather than identifying supporting eigenstates in the full spectrum. Algorithms implementing this idea were developed in~\cite{petrova2025finding, ren2025scarfinder}.
    
    In classical Hamiltonian mechanics, periodic orbits provide an organising framework for phase space, and their stability and bifurcations reveal the transition between integrable and chaotic dynamics. In integrable systems, phase space is foliated by invariant tori on which motion is either quasi-periodic (for incommensurate frequencies) or periodic (for resonant frequencies). Under generic perturbations, KAM theory implies that many nonresonant tori survive while resonant tori are destroyed, producing a mixed phase space consisting of surviving invariant tori, resonant island chains, and chaotic regions. 
    Quantum integrable models exhibit an analogous, but algebraically richer, structure, rooted in the Yang--Baxter equation~\cite{yang1967some, baxter1972partition, baxter1985exactly}. The existence of an $R$-matrix satisfying the Yang--Baxter equation guarantees a one-parameter family of commuting transfer matrices, whose expansion in the spectral parameter generates an extensive tower of mutually commuting conserved charges $Q_n$. The Hamiltonian itself appears as one charge in this tower, and the simultaneous eigenstates of all $Q_n$ are constructed exactly by the Bethe ansatz: the many-body wave function is given as a superposition of plane waves with momenta $k_j$ fixed by elastic two-body scattering, reducing the $N$-body problem to a set of coupled algebraic equations~\cite{karbach1998introduction1, karbach1998introduction2, franchini2017introduction}. Every eigenstate is therefore labelled by its full set of Bethe momenta $\{k_j\}$, and the system is intrinsically non-thermal in the Gibbs sense, with thermalisation replaced by relaxation to a generalised Gibbs ensemble fixed by all conserved charges~\cite{rigol2007relaxation, ilievski2015complete}. This highly constrained structure makes the notion of a quantum scar non-trivial in integrable models: rather than asking whether atypical eigenstates exist against a featureless thermal background, one must ask whether the Bethe-ansatz structure itself can organise or protect special towers of states, and whether the periodic-orbit picture that works so well in chaotic systems carries over to this setting.
    
    In this context, the connection between quantum many-body scars and integrability has attracted growing attention, with existing examples suggesting that integrability can support special eigenstate structures through mechanisms quite distinct from those in chaotic systems. Exact scar towers arise at integrable points from special algebraic structures: the $\eta$-pairing tower of the Hubbard model~\cite{yang1989eta, mark2020eta, moudgalya2020eta} is protected by an $\eta$-$SU(2)$ symmetry. At the same time, the AKLT chain supports an exact scar tower built from the repeated action of a magnon creation operator on a reference state~\cite{moudgalya2018entanglement}. In the XXZ chain, the phantom Bethe states~\cite{popkov2021phantom} obtained from stacked-string excitations at $k=\pi$ 
    form degenerate multiplets. 
    Superpositions of these multiplets result in 
    long-lived spin-helix states, directly observed in cold-atom realisations of the chain~\cite{jepsen2022long}. These states are part of a wider family of exact product-state scars present throughout the XYZ chain, whose dynamical stability under generic perturbations has recently been characterised \cite{bhowmick2025asymmetric, bhowmick2025granovskii}. The loop-$\mathfrak{sl}_2$ symmetry at root-of-unity anisotropy~\cite{deguchi2001sl2, fabricius2001completing, braak2001spectrum, hou2024rational} provides a further example of an integrable algebraic structure organising large degenerate towers beyond the standard Bethe catalogue. 
    
    Several of these structures survive the breaking of integrability: the spin-helix states seed scar dynamics in the non-integrable regime~\cite{jepsen2022long}, and the quantum group $U_q(sl_2)$ can remain as a global symmetry even when integrability is explicitly broken~\cite{gorbenko2025chaos}. Their persistence beyond integrability suggests that the underlying organising principles may be more general than integrability itself. At the same time, integrable models provide direct analytical access to the organisation of the spectrum and eigenstates, a level of microscopic control that is largely absent in generic chaotic systems. This raises the general question of whether the periodic-orbit paradigm of quantum scarring can be understood in terms of the algebraic and geometric structures revealed by integrability. 
    
    To address this question, we apply the periodic-orbit framework to the integrable XYZ and XXZ spin-$1/2$ chains. We extend the algorithm of Ref.~\cite{petrova2025finding} to enforce energy conservation, allowing us to search for orbits at specific energies and track them across energy shells. This reveals how the set of supporting eigenstates changes as the energy of the scar is varied~(Sec.~\ref{sec:scars}). By deforming the model to the XXZ point, we can explicitly resolve the supporting eigenstates in terms of Bethe quantum numbers. We find that these states share a string core at $\mathrm{Re}(k)=\pi$, closely related to the phantom Bethe states of Ref.~\cite{popkov2021phantom}, and are dressed by zero-momentum pairs of magnons (Sec.~\ref{sec:bethe}). This construction allows us to reconstruct both the scarred eigenstates and the associated towers directly from the Bethe ansatz, providing a microscopic realisation of scar phenomenology within an integrable setting.
    
    To probe the stability of this structure beyond integrability, we add a transverse field, breaking both $U(1)$ symmetry and integrability while preserving the orbit; we then compare the orbits in the integrable and non-integrable regimes in the thermodynamic limit (Sec.~\ref{sec:int_b}). These results establish a direct connection between the periodic-orbit picture of quantum scarring and the algebraic structure of integrable models, and demonstrate that key features of scar dynamics—normally identified numerically in non-integrable systems such as the PXP model—can be understood analytically in terms of Bethe-ansatz data, with broader implications discussed in Sec.~\ref{sec:discussion}.
    
    Our results suggest extending the definition of scarred eigenstates to an
    integrable setting, where the conventional definition in terms of a weak ETH violation is unavailable. Our work allows us to define \emph{scarred families of eigenstates} dynamically: a sub-extensive, approximately
    equidistant set of eigenstates that carries an 
    $O(1)$ fraction of the weight of a low-bond-dimension periodic orbit with small leakage. The anomalously low entanglement of these eigenstates is then a consequence of their
    microscopic structure~\cite{alba2009entanglement}. As we discuss below, in contrast to typical Bethe states, whose smooth distribution of occupied Bethe quantum numbers gives rise to volume-law entanglement, the supporting eigenstates concentrate a macroscopic number of Bethe roots at a single Bethe quantum number, dressed by a discrete set of zero-momentum magnon pairs.

    \section{Periodic orbits and scarred eigenstates in the XYZ spin chain} 
    \label{sec:scars}

    \begin{figure}[t]
         \centering
         \includegraphics[width=0.99\linewidth]{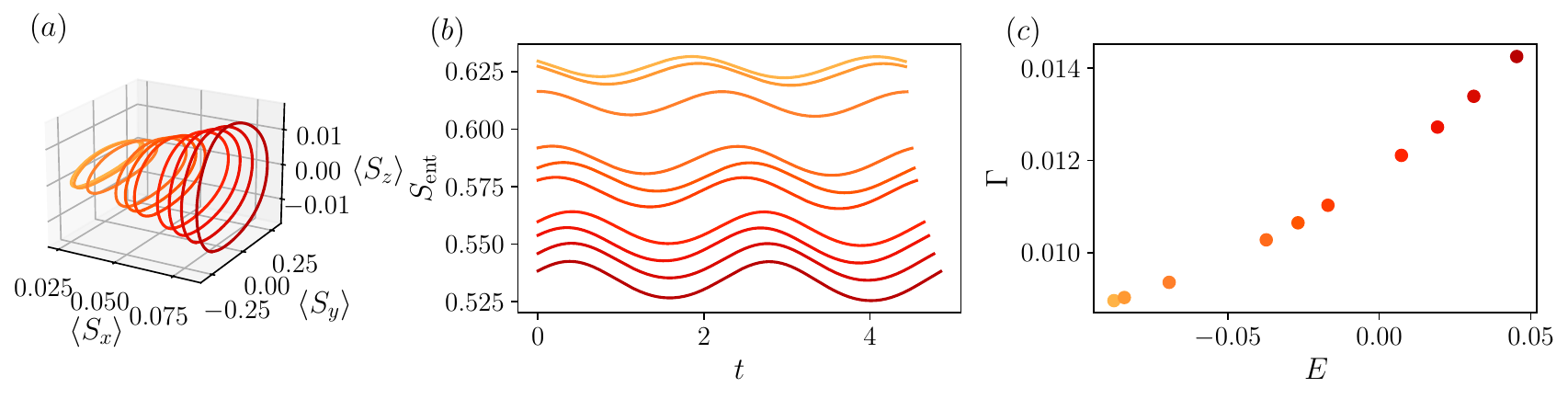}
         \caption{(a) Local observables along the periodic orbit found at bond dimension $\chi = 2$ for the XYZ Hamiltonian ($J_x = 1$, $J_y = 0.8$, $\Delta = -0.5$). By slightly perturbing an orbit and re-optimising it with the gradient descent algorithm, we can continuously track the same orbit across different energy shells. (b) Smooth variation of the entanglement entropy of the orbit across different energy shells. (c) Integrated leakage for different energies. Although the entanglement entropy of the orbit decreases with energy, the leakage increases.}
         \label{fig:traj_xyz}
     \end{figure}

    We consider the XYZ spin-$1/2$ chain described by the Hamiltonian
    \begin{equation}
        \hat{H} = \sum_i J_x S^x_i S^x_{i+1} + J_y S^y_i S^y_{i+1} + \Delta S^z_i S^z_{i+1},
        \label{eq:XYZ}
    \end{equation}
    where $S^{x,y,z}$ are spin-$1/2$ operators, related to the Pauli matrices by 
    $S^{\alpha} = \sigma^{\alpha}/2$. The model is integrable for arbitrary $(J_x, J_y, \Delta)$~\cite{baxter1972partition,baxter1985exactly}. For pairwise distinct couplings $J_x$, $J_y$ and $\Delta$, the internal spin-rotation symmetry is reduced to the discrete $\mathbb{Z}_2 \times \mathbb{Z}_2$ group generated by $\pi$-rotations around the coordinate axes. When two of the couplings coincide, $J_x = J_y \neq \Delta$, the model reduces to the XXZ chain with $U(1)$ symmetry, corresponding to rotations around the $z$-axis. Finally, the fully isotropic point $J_x = J_y = \Delta$ recovers the Heisenberg chain with $SU(2)$ symmetry. 
    
    The XYZ chain is the natural starting point, as it has the fewest symmetries among the cases discussed above, allowing us to study the orbit structure in its most general setting. To identify periodic orbits, we employ the periodic-orbit search algorithm introduced in~\cite{petrova2025finding}, which combines an MPS representation of the wave function with TDVP evolution. The connection to the XXZ chain is restored later in Sec.~\ref{sec:bethe} by continuously tuning $J_y \to J_x$.

    Using 50 random initialisations, we find periodic orbits at bond dimensions $\chi = 2, 3$ and $4$ for the XYZ chain at $J_x = 1$, $J_y = 0.8$, $\Delta = -0.5$.\footnote{The $\chi = 2$ orbit is obtain with a time step $\delta t = 10^{-3}$ in TDVP evolution. For $\chi = 3$ and $\chi = 4$, the variational manifold is substantially larger, and the search becomes correspondingly more expensive. As a result, these orbits are obtained using a larger time step $\delta t = 10^{-2}$.}
    After the post-processing procedure, which filters out eigenstates and local minima, we obtained $8$ orbits that occupy noticeably different mean energies in the bulk yet concentrate their weight on the same small subset of eigenstates of $\hat {H} $, a sparse collection of states distributed throughout the spectrum. The fact that uncorrelated initialisations repeatedly recover the same set of supporting eigenstates suggests that the algorithm identifies a single underlying set of scarred eigenstates and that the same scar structure can support coherent dynamics at several different energies. We note that the variational search is a priori biased towards eigenstates with low entanglement, but the non-trivial observation is that the supporting eigenstates also form an approximately equidistant ladder structure that the manifold itself does not impose and that we confirm at large $L$ through the revivals of local observables in Sec.~\ref{sec:int_b}.

    To study this energy dependence systematically, we would like to follow a single orbit as its energy is varied, rather than relying on randomly sampled orbits. The algorithm in Ref.~\cite{petrova2025finding}, however, does not conserve energy, and the optimisation will typically cause the orbit's energy to drift. We therefore perturb an orbit towards a neighbouring energy shell and modify the algorithm so that, at every subsequent optimisation step, the update is constrained to leave the total energy unchanged to first order. Imposing this constraint removes one direction from the variational tangent space and forces the optimisation to proceed within the energy shell of the perturbed orbit rather than drifting across shells. The derivation of the constrained update, its explicit form, and a comparison of the optimisation behaviour with and without the energy constraint are given in App.~\ref{app:algorithm}.
    
    The resulting smooth one-parameter family at $\chi = 2$ is shown in Fig.~\ref{fig:traj_xyz}(a) and (b) (See Appendix for $\chi = 3,4$ data), and the bipartite entanglement entropy obtained from iTDVP with fixed bond dimension smoothly oscillates along the orbit near a value slightly below $\ln 2$, the maximal entropy possible in a $\chi=2$ MPS. 
    To assess how closely each member of this family approximates the true quantum dynamics, we compute the instantaneous leakage~\cite{ho2019periodic, michailidis2020slow,ljubotina2022optimal}:
    \begin{equation}
        \gamma^2(t) = \left| \partial_t |\psi(t)\rangle + i\hat{H}|\psi(t)\rangle\right|^2,
    \end{equation}
    which we then integrate over one period. We denote the integrated leakage by $\Gamma = \frac{1}{T}\int_0^T \gamma^2(t) dt$ and plot it in Fig.~\ref{fig:traj_xyz}(c). Counter-intuitively, the integrated leakage \emph{grows} with energy even though the entanglement entropy along the orbit \emph{decreases}, indicating that the quality of the projected orbit is not controlled by entanglement alone.

\begin{figure}
        \centering
        \includegraphics[width=0.99\linewidth]{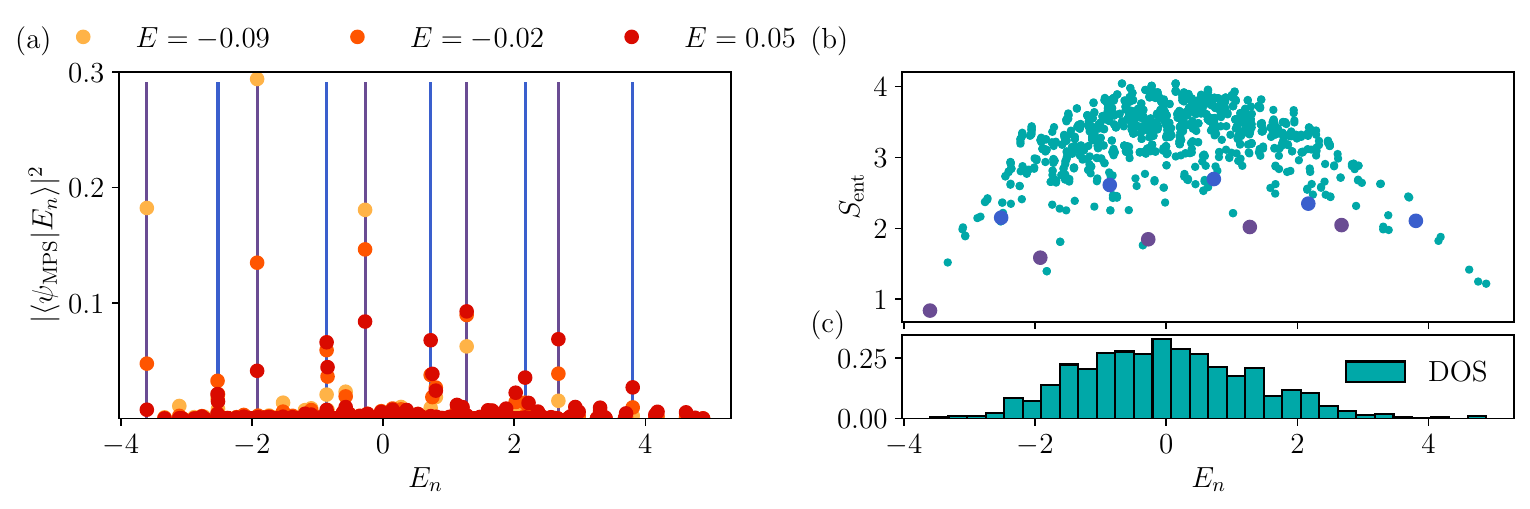}
        \caption{(a) Overlap of the MPS state with eigenstates of the Hamiltonian for $L = 14$. Purple and blue lines correspond to the eigenstates that contribute most to the scars. (b) Entanglement entropy of the eigenstates of the XYZ Hamiltonian. Purple and blue dots correspond to the eigenstates that contribute most to the scars. The two sets contain states that are approximately equidistant, which can give rise to periodic dynamics. As the energy of the scar changes, it slowly shifts from the purple eigenstates to a combination of purple and blue. (c) Density of states showing no atypical behaviour in the spectrum. }
        \label{fig:scars}
    \end{figure}

    To understand the structure of the orbits at different energies, we diagonalise the Hamiltonian on a finite chain of $L = 14$ sites and identify the eigenstates with the largest overlap with the orbit, as shown in Fig.~\ref{fig:scars}. The supporting eigenstates split into two approximately equidistant subsets, marked by violet and blue lines in Fig.~\ref{fig:scars}(a) and by the corresponding dots in Fig.~\ref{fig:scars}(b). Both subsets exhibit anomalous entanglement entropy compared with the bulk of the spectrum. As the scar energy is tuned, the weight carried by the supporting eigenstates shifts smoothly: it is initially confined to the violet subset, is then redistributed among the eigenstates within this subset, and finally spreads across both the violet and blue subsets. The transition between scar branches is therefore not a sharp event but a smooth redistribution of weight within an equidistant tower of supporting states.

    This picture also provides a possible explanation for the counter-intuitive leakage trend in Fig.~\ref{fig:traj_xyz}: as the scar energy increases, the supporting eigenstates become more entangled, while the variational manifold remains restricted to $\chi=2$. The exact Schr\"odinger time-evolution direction may therefore acquire a larger component outside the variational tangent space, resulting in increased leakage. The simultaneous decrease in the entanglement of the orbit itself may reflect the restriction imposed by the fixed-bond-dimension MPS manifold: as the entanglement of the supporting eigenstates increases, their representation within the $\chi=2$ manifold requires progressively stronger compression. To understand the microscopic origin of these approximately equidistant towers and their changing entanglement structure, we now turn to the integrable XXZ point, where the Bethe ansatz allows us to resolve the supporting eigenstates explicitly.
    
    \section{Bethe ansatz structure of the scars}
    \label{sec:bethe} 

    \begin{figure}
        \centering
        \includegraphics[width=0.6\linewidth]{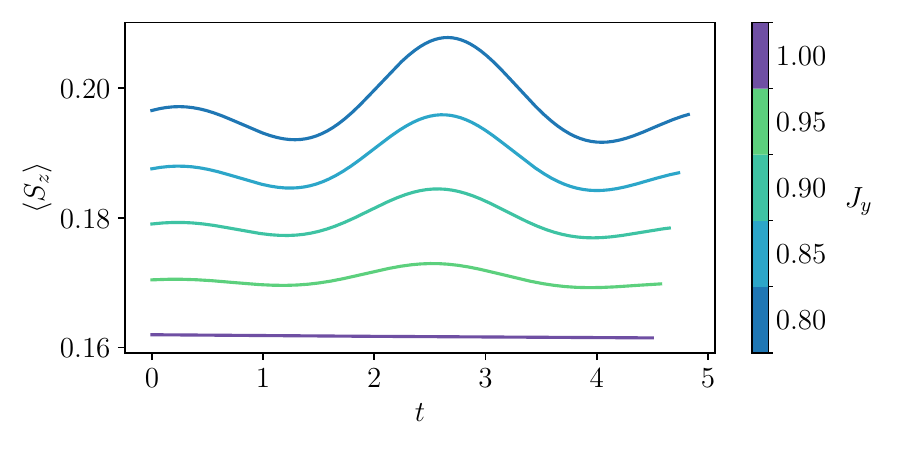}
        \caption{Evolution of the magnetisation density $\langle S_z(t)\rangle$ along the orbit as the Hamiltonian is deformed from the XXZ point at $\Delta/J=-1.1$ towards the XYZ regime by varying $J_y$. At the XXZ point, the $U(1)$ symmetry implies conservation of the total magnetisation. Although this conservation need not be reproduced exactly by the projected TDVP dynamics, we observe that the magnetisation remains constant at the XXZ point and becomes time-dependent as $J_y$ is tuned away from $J_x$, thereby breaking the $U(1)$ symmetry.}
        \label{fig:XYZ_XXZ}
    \end{figure}

    Although the XYZ model is integrable, its Bethe-ansatz solution involves elliptic functions and does not admit a transparent quasiparticle interpretation for general parameter values. The XXZ point, by contrast, is solved by the coordinate Bethe ansatz, with a well-developed picture of magnons and bound states described by string solutions of different lengths. To make the orbit accessible to such an analysis, we deform the model to the XXZ point in two stages: first, we tune $\Delta$ to $-1.1$, placing the system in the ferromagnetic regime. The orbit is tracked continuously along this first stage so that the family analysed below is smoothly connected to the orbits of Sec.~\ref{sec:scars}. We then smoothly tune $J_y$ towards $J \equiv J_x$. Figure~\ref{fig:XYZ_XXZ} shows the $\chi = 2$ orbit along the second part of this deformation. The periodic orbit persists as a continuous family, and, at the XXZ point, the $U(1)$ symmetry is restored and the magnetisation $\langle S^z(t)\rangle$ becomes constant along the orbit. The $\chi=3$ case is more delicate. Along the same deformation, the optimisation slows down considerably, with very small gradients in some tangent directions and finite gradients in others. Reaching a fully periodic orbit at the XXZ point is therefore prohibitively expensive, and we stop the optimisation at a partially converged orbit. Its supporting eigenstates and scar structure are nonetheless clearly identifiable, allowing us to analyse their characteristic features below using this partially converged orbit. Our goal in this section is to use the Bethe-ansatz structure to explain, at a microscopic level, the equidistant towers of supporting eigenstates observed in Sec.~\ref{sec:scars}.

    \subsection{Numerical solution of the Bethe equations}
    \label{sec:bethe-numerics}
    
    In the ferromagnetic regime $\Delta/J < -1$, the ground state of the XXZ chain is doubly degenerate, corresponding to the fully polarised states $|\!\uparrow\uparrow\cdots\uparrow\rangle$ and $|\!\downarrow\downarrow\cdots\downarrow\rangle$. Excitations above either ground state are described by the coordinate Bethe ansatz~\cite{karbach1998introduction1, franchini2017introduction}. In its quasiparticle interpretation, a single flipped spin propagates as a magnon, a one-particle excitation of momentum $k$, while configurations with several flipped spins are described as states of multiple magnons that interact through the anisotropy $\Delta/J$ and can either scatter off one another or form bound complexes (string solutions). The Bethe ansatz expresses every $M$-magnon eigenstate as a superposition of plane waves with momenta $\{k_j\}_{j=1}^{M}$. The amplitudes are fixed by elastic two-body scattering, while the momenta are quantised by imposing periodic boundary conditions on a chain of length $L$. The resulting Bethe equations are conditions on the magnon momenta $\{k_j\}$ alone, with the two-body interaction entering only through the scattering phase between pairs of magnons. Their solutions are labelled by a set of integer Bethe quantum numbers $\{I_j\}_{j=1}^{M}$ that index the eigenstates. We solve these equations directly in momentum form. The two-magnon scattering phase is~\cite{karbach1998introduction1}
    \begin{equation}
        e^{i\theta(k_1,k_2)} 
        = -\,\frac{e^{i(k_1+k_2)} + 1 - 2\Delta J^{-1}\, e^{ik_1}} 
                  {e^{i(k_1+k_2)} + 1 - 2\Delta J^{-1}\, e^{ik_2}}, 
        \label{eq:scattering}
    \end{equation}
    and the logarithmic Bethe equations read
    \begin{equation}
        L\,k_j \;=\; 2\pi I_j \,+\, \sum_{l\neq j} \theta(k_j, k_l),
        \qquad j = 1,\ldots,M.
        \label{eq:loBAE}
    \end{equation}
    
    With the Hamiltonian convention of Eq.~\eqref{eq:XYZ}, the energy of a Bethe state with $M$ magnons of momenta $\{k_j\}$ is
    \begin{equation}
        \begin{gathered}
        E \;=\; E_0 \;+\; \sum_{j=1}^{M} \varepsilon(k_j), 
        \qquad
        \varepsilon(k) = J\cos k - \Delta,
        \\
        E_0 \;=\; \frac{\Delta L}{4},
        \end{gathered}
        \label{eq:bethe-energy}
    \end{equation}
    where $E_0$ is the polarised reference-state energy and $\varepsilon(k)$ is the XXZ single-magnon dispersion. In the ferromagnetic regime $\Delta/J < -1$ considered here, $\varepsilon(k) > 0$ for all $k$, so each magnon adds energy above the polarised ground state. The total momentum is fixed by the quantum numbers:
    \begin{equation}
        K \;=\; \sum_{j=1}^M k_j \;=\; \frac{2\pi}{L} \sum_{j=1}^M I_j 
        \pmod{2\pi}.
        \label{eq:total-momentum}
    \end{equation}
    For further details on solving the Bethe equations and constructing the corresponding eigenstates, see App.~\ref{app:Bethe}.

    \begin{figure}
        \centering
        \includegraphics[width=1\linewidth]{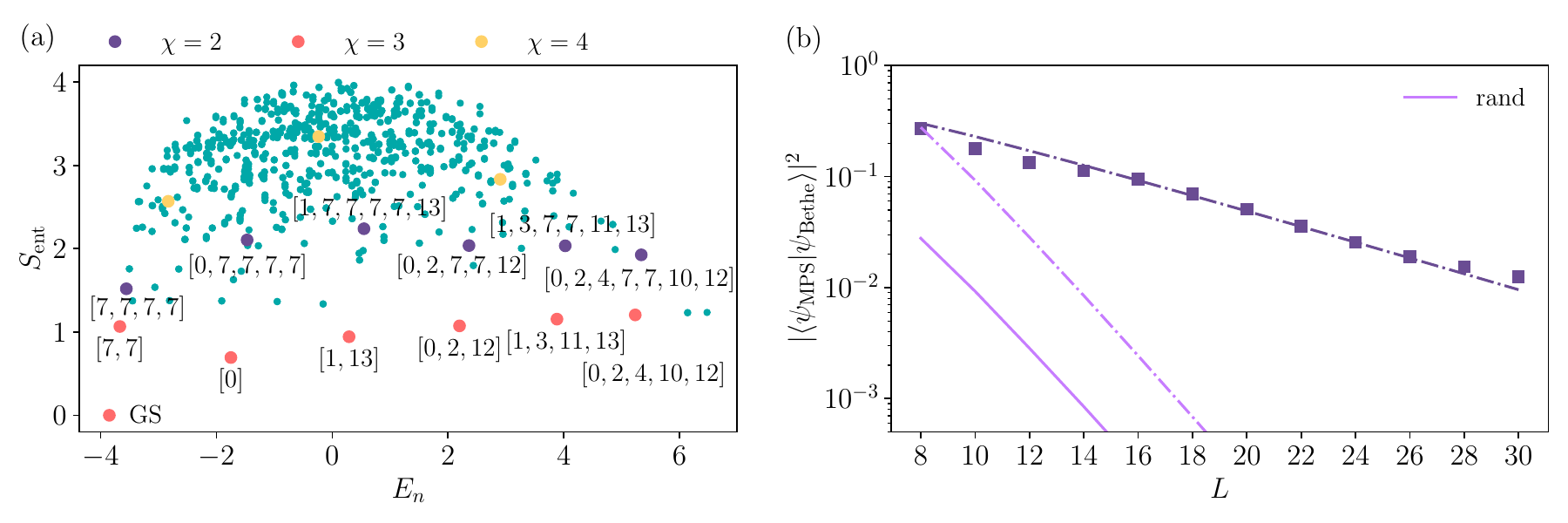}
        \caption{(a) Entanglement entropy of eigenstates of the XXZ chain at $L=14$, $\Delta=-1.1$, with dots marking the eigenstates that contribute most to the scars at $E/L=0.046$ with $\chi=2,3,4$. The numbers in brackets are the Bethe quantum numbers $\{I_j\}$ of the corresponding eigenstate. (b) The \emph{maximum} overlap with the scar orbit as a function of the system size $L$ (XXZ chain, $\Delta = -1.1$, $\chi=2$). The supporting eigenstate for each $L$ decomposes into a ``core'' of $2n$ identical roots at $I = L/2$  together with zero-momentum dressing units: symmetric pairs $\{a, L-a\}$ and, when present, a lone zero-momentum magnon at $I=0$. Although the maximum overlap decays exponentially with system size, its decay is substantially slower than that of the overlap with random states in the zero-momentum sector, suggesting that the scar structure remains identifiable as the system size increases. The energy density of the maximum-overlap eigenstate varies with $L$, but the relevant eigenstate remains in the bulk of the many-body spectrum.}
        \label{fig:bethe-id}
    \end{figure}
    
    \subsection{Bethe structure of bond dimension 2 and 3 scars}
    
\begin{figure}
        \centering
        \includegraphics[width=0.7\linewidth]{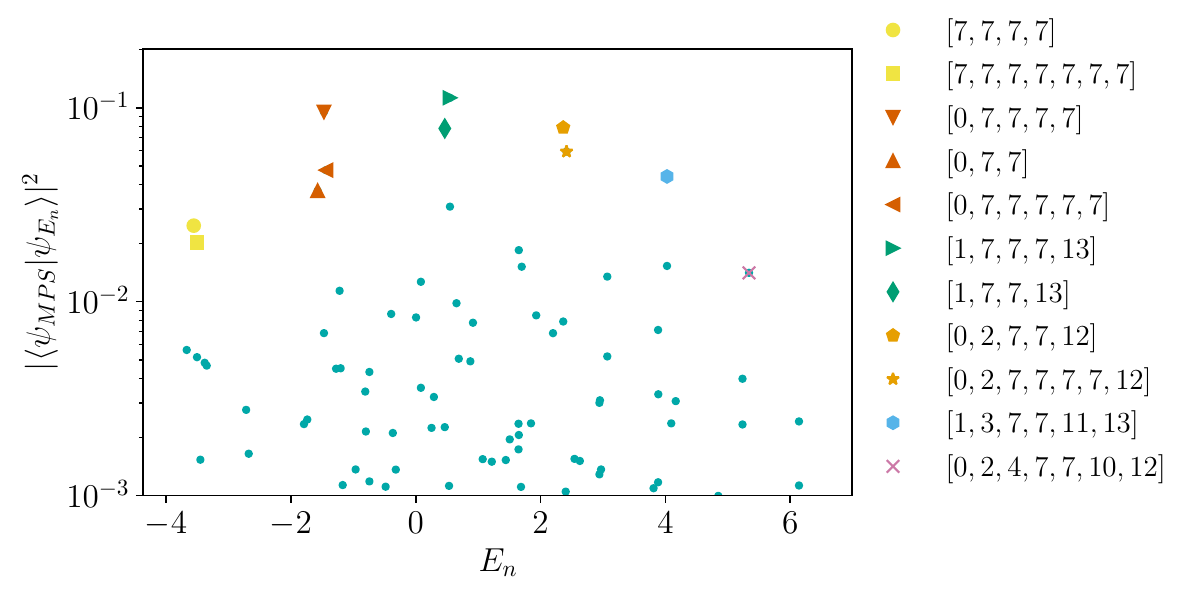}
        \caption{Dependence of the overlap between the scar state and the Hamiltonian eigenstates on the size of the 
         $I=L/2$ core in the Bethe quantum-number configuration at $L=14$, $\Delta=-1.1$. Each colour corresponds to a distinct dressing pattern. Within each pattern, the states differ only in the number of copies of $I=L/2$ and have comparable energies and overlaps with the $\chi=2$ MPS orbit, thereby forming a tower. Adding a pair of roots with $I=L/2$ to the core shifts the energy by an amount much smaller than the spacing between the adjacent towers.}
        \label{fig:core_degeneracy_L14}
    \end{figure}

    Throughout this section, we restrict our analysis to chains of even length $L$. This choice simplifies the Bethe-state classification while retaining a sequence of system sizes along which the thermodynamic limit can be taken. For both the $\chi=2$ and $\chi=3$ orbits, we identify the Bethe quantum numbers $\{I_j\}$ of the eigenstates that carry most of the orbit's weight. The two cases share the same building blocks--pairs of excitations with zero net momentum--which generate an approximately equidistant set of eigenstates that we refer to throughout this paper as a \emph{ladder}. The difference between the two cases is that the $\chi=2$ ladder additionally contains a macroscopic bound-state core, which is absent for $\chi=3$. Figure~\ref{fig:bethe-id}(a) shows the entanglement entropy of the $L=14$ XXZ spectrum in the zero-momentum sector, with the dominant supporting eigenstates of each orbit highlighted and their Bethe quantum numbers annotated, while Fig.~\ref{fig:core_degeneracy_L14} displays the overlap distribution of the $\chi=2$ orbit over the XXZ eigenstates. Two notable features emerge. First, the dominant support is concentrated on an approximately equidistant ladder of eigenstates, reflecting the characteristic scar structure. Second, the overlap structure reveals that it is not a single sequence of equidistant eigenstates that supports the orbit. Instead, each state in the ladder is accompanied by a \emph{tower} of nearly degenerate eigenstates. A complete analysis of the overlap distribution is presented in App.~\ref{app:xxz}. We focus on the richer $\chi=2$ case in the remainder of this subsection and return to $\chi=3$ towards the end.

    For the $\chi=2$ orbit, the core consists of $2m$ magnons sharing the same Bethe quantum number $I_j = L/2$ for even $L$. Solving the Bethe equations for this configuration yields complex momenta: the magnons sharing the same Bethe quantum number form a bound state with $\operatorname{Re}(k_j)=\pi$, with the momenta occurring in complex-conjugate pairs.
    The approximately equidistant ladder of eigenstates supporting the orbit is then obtained by dressing this core, with the form of the dressing differing slightly between the even- and odd-$M$ sectors:
    \begin{enumerate}
        \item \emph{Even magnon number $M$:} A pair of magnons with Bethe quantum numbers $\{a,L-a\}$, where $1\leq a<L/2$ and $a$ is an odd integer.
        \item \emph{Odd magnon number $M$:} A single magnon with Bethe quantum number $I=0$, together with a pair of magnons with Bethe quantum numbers $\{a,L-a\}$, where $1\leq a<L/2$ and $a$ is an even integer.
    \end{enumerate}
    Each dressing pair is individually momentum-neutral, and the core contributes a total momentum $2m\pi \equiv 0 \pmod{2\pi}$. 
    Therefore, all scarred eigenstates identified through their overlap with the orbit belong to the $K=0$ sector, as expected for the one-site translation-invariant MPS orbit.
    
    The nearly degenerate eigenstates accompanying each state in the ladder originate from a distinct structure in the Bethe-root configurations. States that share a common dressing but differ in the number of $I = L/2$ core roots cluster within a narrow energy window and carry comparable overlaps with the MPS orbits, as seen in Fig.~\ref{fig:core_degeneracy_L14}. 
    An analogous mechanism was studied in~\cite{popkov2021phantom}, where stacked-string excitations at $k = \pi$ produce exact energy degeneracies and underlie the so-called phantom Bethe states. In the ferromagnetic regime considered here, the supporting eigenstates of the scar exhibit the same Bethe-root structure: a stacked string at the Bethe quantum number $I=L/2$ with zero-momentum dressing. However, the precise energetic features of the phantom states do not carry over directly. Instead of producing an exact degeneracy, variations in the size of the core generate a narrow cluster of energies, giving rise to the tower structures depicted in Fig.~\ref{fig:core_degeneracy_L14} (see App.~\ref{app:xxz} for a larger system size). 
    
    The structure of the Bethe equations~\eqref{eq:loBAE} explains why each zero-momentum dressing pair shifts the energy by approximately the same amount. With $S(k_1,k_2)=e^{i\theta(k_1,k_2)}$, the scattering contribution on the right-hand side of Eq.~\eqref{eq:loBAE} is the sum of the phase shifts experienced by magnon $j$, whose exponential is $\prod_{\ell\neq j}S(k_j,k_\ell)$. The equation for $k_j$ therefore decouples from the remaining Bethe equations whenever the total scattering contribution is independent of the other momenta. This occurs at $k=0$. For a two-string core, direct calculation gives $S(0,\pi+i\im k_b)S(0,\pi-i\im k_b)=1$, independently of $\Delta$ and $\im k_b$. Similarly, $S(0,k_a)S(0,2\pi-k_a)=1$ for any symmetric pair $\{k_a,2\pi-k_a\}$ and arbitrary $\Delta$. The exactly zero-momentum magnon therefore decouples from the core and from every symmetric pair, contributing a fixed energy independent of the remaining Bethe-root configuration. For a symmetric pair of Bethe quantum numbers $\{a,L-a\}$ with small but finite momentum $k_a$, the scattering of one dressing magnon with the core gives $S(k_a,\pi+ib)S(k_a,\pi-ib)=1+\mathcal{O}(k_a)$, while its scattering with another dressing pair similarly gives $S(k_a,2\pi-k_c)S(k_a,k_c)=1+\mathcal{O}(k_a)$. For fixed $a$, $k_a=\mathcal{O}(L^{-1})\to0$ as $L\to\infty$, so this residual coupling is suppressed by $1/L$ and the dressing-ladder spacing approaches its $k=0$ limit. 
    
    Interestingly, we observe that the equidistant spacing persists even at higher energy densities, beyond the $E/L\approx E_0/L$ regime in which the above argument is expected to apply. Each dressing unit therefore contributes approximately the same energy, producing the equidistant spacing shown in Fig.~\ref{fig:bethe-id}. The $\chi=3$ orbit in Fig.~\ref{fig:bethe-id} exhibits the same structure: its supporting eigenstates follow the $\chi=2$ dressing family but without the $\{L/2,\ldots,L/2\}$ core. Accordingly, the same equidistant ladder reappears.
    
    Figure~\ref{fig:bethe-id}(b) shows the scaling of the maximum overlap between the scarred orbit and its dominant supporting eigenstate with system size (see App.~\ref{app:xxz} for the exact Bethe quantum-number structure of these states). Equipped with the Bethe quantum-number structure of the relevant states, we construct the dominant supporting eigenstate directly from the Bethe ansatz rather than using exact diagonalisation. For each $L$, we solve the Bethe equations~\eqref{eq:loBAE} for the quantum-number pattern described above and construct the coordinate Bethe-ansatz wave function in the zero-momentum sector and the corresponding magnetisation sector. We then evaluate the overlap of this eigenstate with the MPS orbit. This procedure allows us to determine the scaling beyond the system sizes accessible by exact diagonalisation, even after accounting for all symmetries.

    We observe that the overlap decays exponentially with $L$, consistent with the approximation of a many-body eigenstate by a fixed-bond-dimension MPS. Its decay rate is nevertheless substantially lower than that obtained for Gaussian random states within the zero-momentum sector. This slower decay indicates that the scarred orbit retains anomalously large overlap with a selected subset of eigenstates rather than behaving as a random state in the same symmetry sector. The ladder-and-tower overlap structure also persists at $L=30$, as shown in Fig.~\ref{fig:scar30} of App.~\ref{app:xxz}, well beyond the system sizes accessible by exact diagonalisation.

    As shown in Sec.~\ref{sec:scars}, the scar trajectories in the XYZ model are smooth deformations of the Bethe-tower structure analysed in this section. This indicates that the scar structure is stable under integrability-preserving deformations. What remains unclear is whether this structure persists once integrability is broken.
    
    \section{Integrability breaking}
    \label{sec:int_b}
    
    \begin{figure*}
        \centering
        \includegraphics[width=1.\linewidth]{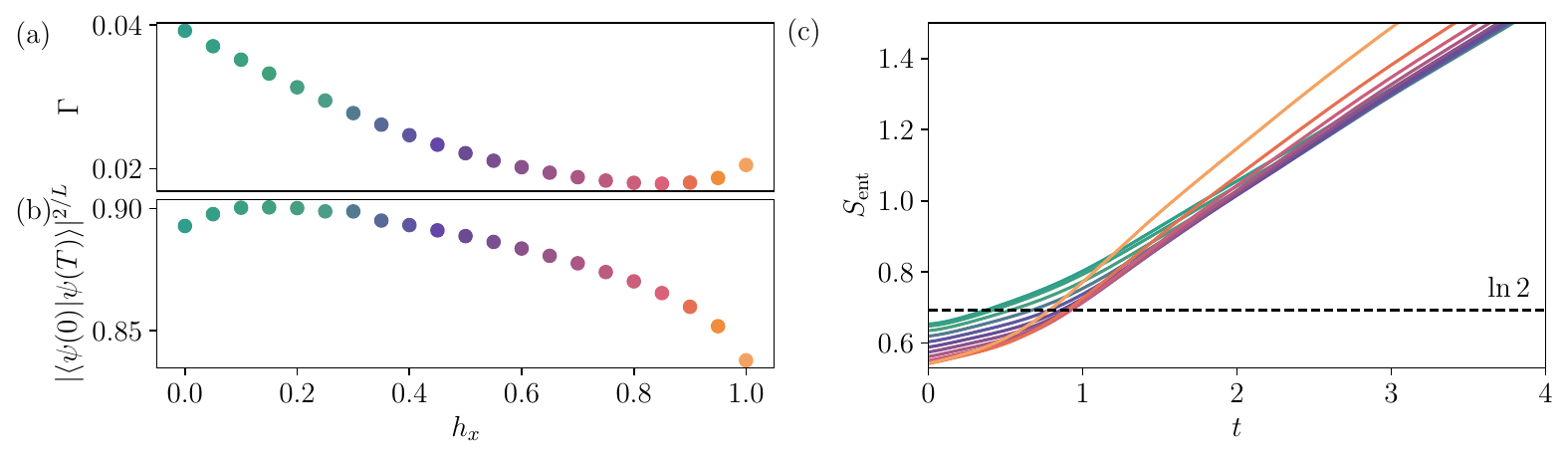}
        \caption{
        (a) The integrated leakage $\Gamma$ over one period for different orbits decreases with increasing field strength, indicating that the integrable point ($h_x = 0$) does not necessarily correspond to minimum leakage. (b) Fidelity per site after one period for $L=200$ and $\chi=300$, obtained from TEBD simulations of the quantum dynamics. Despite the decrease in leakage, revivals worsen as integrability is broken. (c) Entanglement growth in the TEBD evolution; colours encode the value of $h_x$ as in panels (a) and (b). For $h_x$ close to $1$, the entanglement remains below $\log 2$ for the longest time, suggesting that the early-time dynamics can be captured by a $\chi=2$ MPS and explaining the small leakage. At later times, the entanglement grows faster for the non-integrable orbits, consistent with the reduced fidelity in panel (b).} 
        \label{fig:int_b}
    \end{figure*}
    
    We probe the robustness of the periodic orbit to integrability-breaking perturbations by adding a uniform transverse field to the XXZ Hamiltonian:
    \begin{equation}
        \hat{H} = \hat{H}_{\rm XXZ} + h_x \sum_i S^x_i.
        \label{eq:Hbreak}
    \end{equation}
    We consider the one-parameter family of orbits obtained by running the algorithm of Sec.~\ref{sec:scars} for $h_x \in [-0.5, 1]$, seeded by the integrable orbit at $h_x = 0$. Importantly, the transverse field in the $x$ direction simultaneously breaks the $U(1)$ symmetry of the XXZ point and the integrability of the model \cite{shiraishi2019proof}. Therefore, any surviving orbit cannot be protected by either the $U(1)$ symmetry of the XXZ point or the integrability of the model.
    
    Figure~\ref{fig:int_b}(a) shows the integrated leakage $\Gamma$ over one period as a function of $h_x$ for the family of orbits defined by the Hamiltonian~\eqref{eq:Hbreak}. For negative fields, the leakage increases, so we follow the family towards positive fields. The leakage then decreases monotonically to a minimum at $h_x\approx0.9$, deep in the non-integrable regime, before increasing again. In particular, $\Gamma$ is not minimised at the integrable point $h_x = 0$. This asymmetry is not in conflict with the unitary equivalence of $\hat{H}(h_x)$ and $\hat{H}(-h_x)$ under a $\pi$ rotation about the $z$ axis. Because the orbit at $h_x=0$ is not invariant under this rotation, its direct continuations towards positive and negative $h_x$ need not be related by the symmetry; instead, rotating the initial orbit generates a mirror family with the reflected leakage profile $\Gamma(-h_x)$.
    
    The integrated leakage $\Gamma$ quantifies the failure of the $\chi=2$ manifold to accommodate the exact Schr\"odinger evolution. A small $\Gamma$ at $h_x\approx0.9$ therefore indicates that the exact Schr\"odinger time-evolution direction along the deformed orbit is better captured by the tangent space of the $\chi=2$ MPS manifold than along its integrable counterpart. This trend is accompanied by larger entanglement entropy and a larger magnitude of the subleading eigenvalue of the MPS transfer matrix for the integrable orbit. It does \emph{not} by itself establish that integrability is irrelevant to the scarred dynamics, which would require a manifold-independent diagnostic. We therefore complement the leakage with large-bond-dimension quantum dynamics. Running TEBD simulations at $L = 200$, $\chi = 300$, we track the revival fidelity per site $F_{PS} = |\langle \psi(t)|\psi(0)\rangle|^{2/L}$ and the entanglement entropy (Fig.~\ref{fig:int_b}(b,c)), revealing that the least-leaky orbit at $h_x \approx 0.9$ is not the one with the best fidelity revival. We return to the discrepancy between the leakage and fidelity diagnostics at the end of this section.

    \begin{figure}
        \centering
        \includegraphics[width=.99\linewidth]{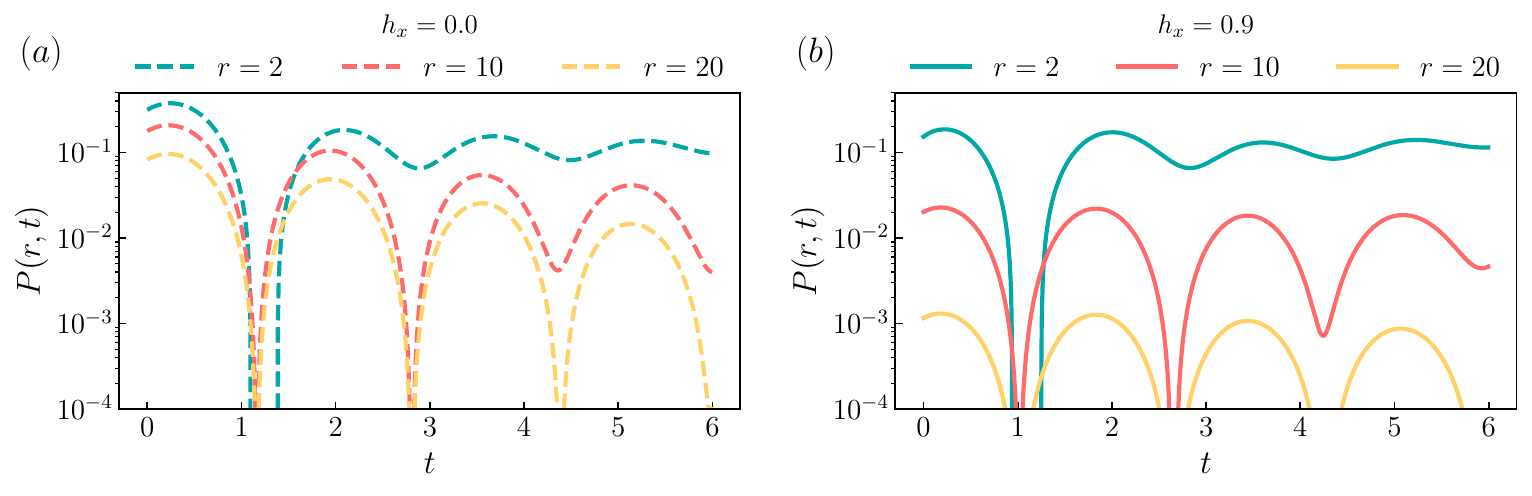}
        \caption{Dynamics of the connected two-point correlator $P(r,t)$, defined in Eq.~\eqref{eq:Prt}, for the orbits at $h_x=0$ and $h_x=0.9$, shown in panels (a) and (b), respectively. The $h_x=0$ trajectory exhibits correlations of larger magnitude at bigger $r$, whereas the $h_x=0.9$ trajectory exhibits an exponential decay with $r$ and a slower decay in time.}
        \label{fig:revival1}
    \end{figure}

    To characterise the correlations in the TEBD dynamics, we consider the connected two-point correlator
    \begin{equation}
        P(r,t) = \frac{1}{L}\sum_i \langle \psi(t)| P_i P_{i+r}|\psi(t)\rangle_c ,
        \label{eq:Prt}
    \end{equation}
    with $P_i = (1 - \sigma^z_i)/2$ being the local projector onto the down-spin state. Figure~\ref{fig:revival1} provides a quantitative test of the Bethe picture of Sec.~\ref{sec:bethe} at $L=200$.
    Specifically, the Bethe analysis predicts a tower of scar states with nearly uniform energy spacing, implying coherent oscillations. We now compare this prediction with the TEBD dynamics. Here, two observations stand out. First, $P(r,t)$ for $r=2,10,20$ exhibits clear and sustained revivals throughout the interval $t\in[0,5]$ in both the integrable and non-integrable regimes, confirming that the orbit is not a finite-size artefact. Second, in addition to the recurrence at $t=T\approx3.3$, which coincides with the fidelity revival, we observe another recurrence near $t\approx T/2$. The primary revival is consistent with the period expected from the scar tower spacing, $T = 2\pi/\Delta E \approx 2\pi/1.9 \approx 3.3 $. The half-period revival originates from a property of the observable we measure. Since $P_i P_{i+r}$ commutes with the total magnetisation, at the XXZ point, its matrix elements vanish between eigenstates in different magnetisation sectors $M$. The observable therefore acts as a filter: among all pairs of ladder states, only those belonging to the same magnetisation sector contribute to its oscillations. This feature is visible in Fig.~\ref{fig:bethe-id} of the previous section. The states $[0,7,7,7,7]$ and $[0,2,7,7,12]$ belong to the same magnetisation sector, as do the states $[1,7,7,7,7,13]$ and $[1,3,7,7,11,13]$. Within each pair, the states are separated by $2\Delta E$, twice the main ladder spacing, producing the period $T/2$ in the revival of the observable. The same pattern is seen in the $L = 30$ scar structure in App.~\ref{app:xxz}. In the integrability-broken regime, there is no $U(1)$ symmetry, but this feature remains a fingerprint of the integrable model from which we started. 

    We now turn to the differences between the two orbits. First, for the orbit at $h_x=0.9$, the peak amplitude of $P(r,t)$ after one period remains close to its initial value, in sharp contrast to the integrable orbit, for which the correlators decay noticeably over the same interval. Second, when integrability is broken, the correlations become more strongly localised in space: $P(r=10,t)$ and $P(r=20,t)$ are significantly smaller than $P(r=2,t)$. This is consistent with the larger magnitude of the subleading transfer-matrix eigenvalue for the integrable orbit, which corresponds to a longer correlation length for connected correlation functions.
     
    Finally, we return to the disagreement between the leakage and fidelity diagnostics, which stems from their sensitivity to different aspects of the dynamics. The integrated leakage $\Gamma$ quantifies how well the exact Schr\"odinger evolution along the orbit is accommodated by the tangent space of the $\chi=2$ manifold. The fidelity per site, by contrast, compares the full many-body wave functions and is therefore sensitive to changes distributed across the entire system. Away from the integrable point, the weaker late-time fidelity revivals are accompanied by faster entanglement growth, as shown in Fig.~\ref{fig:int_b}(c). Local correlators are less directly sensitive to changes in the wave function distributed over distant degrees of freedom and are also experimentally accessible. The apparent deterioration of the revivals away from integrability is therefore specific to the global fidelity diagnostic. By contrast, the local correlators exhibit stronger revivals for the orbit at $h_x\approx0.9$ than for the integrable orbit.

    \section{Discussion} 
    \label{sec:discussion} 
    
    In this work, we consider quantum many-body scars as periodic orbits of low-dimensional variational dynamics in a setting where exact tools are available, and obtain several key findings. We modify the orbit-finding algorithm of Ref.~\cite{petrova2025finding} to enforce energy conservation to first order. This allows us to find a family of periodic orbits in the XYZ chain, a model with no continuous symmetry, and track them continuously across energy shells. In doing so, we uncover the mechanism by which an orbit shifts between its supporting scarred eigenstates as its energy is varied. Upon deforming the model to the XXZ point, a microscopic picture emerges: the relevant eigenstates share a common $\{L/2,\ldots, L/2\}$ stacked-string core at $\operatorname{Re}(k)=\pi$ and differ in their zero-momentum dressing. Furthermore, by varying the number of magnons in the core and the dressing composition, we can reconstruct the ladder-and-tower structure of the scarred eigenstates. Breaking integrability with a transverse field does not destroy the orbit. Instead, the integrated leakage is minimised at $h_x\approx0.9$, deep in the non-integrable regime. Nevertheless, direct TEBD simulations at large $L$ show that the fidelity revivals are strongest at the integrable point, demonstrating that minimising $\Gamma$ does not necessarily optimise the global revival fidelity.
    
    The microscopic structure of the scar-supporting eigenstates naturally explains the associated towers of nearly degenerate states revealed by their overlaps with the $\chi=2$ scarred orbit. The clustering within each tower originates from the size of the bound state formed by magnons with $\operatorname{Re}(k)\simeq\pi$ and can be understood as a near-cancellation of the energy contributed by each additional pair of core magnons. This mechanism is closely related to the phantom-excitation construction of Ref.~\cite{popkov2021phantom}. In the easy-plane regime, the phantom excitations carry exactly zero energy at special anisotropies. In our case, $\Delta<-1$, the stacked-string core can be viewed as an analogue of this phantom-root mechanism, although, in the present finite periodic system, the cancellation is only approximate rather than exact.

    The interpretation in terms of Bethe quantum numbers also clarifies the origin of the equidistant spacing between the scar towers, which forms the ladder structure. Because the Bethe equations approximately decouple the dressing magnons from the $\pi$-momentum core, the scattering of each dressing pair with the core and with the other symmetric pairs becomes asymptotically trivial, and each pair therefore contributes an approximately fixed energy increment independent of the rest of the state.
    
    This Bethe-ansatz mechanism supports the definition of scarred families of eigenstates proposed in the introduction: the dynamically identified family is distinguished microscopically by the condensation of an extensive number of Bethe roots at a single Bethe quantum number, in contrast to the smooth thermodynamic distribution of occupied Bethe quantum numbers in typical Bethe states. This mechanism also suggests a broader analytical approach to quantum many-body scars in integrable models.

    In the present work, the microscopic identification of the supporting eigenstates was carried out at the $U(1)$-symmetric XXZ point, where the coordinate Bethe ansatz provides explicit eigenfunctions that can be contracted with the MPS orbit. The same procedure extends naturally to other $U(1)$-symmetric chains solvable by the coordinate Bethe ansatz, such as the higher-spin Zamolodchikov--Fateev family~\cite{zamolodchikov1980model,kulish1981yang}, in which the Bethe structure is more complex, but the eigenstates remain analytically accessible. The Hubbard model is particularly interesting in this context because Yang's $\eta$-pairing states form an exact non-thermal tower generated by a spectrum-generating algebra~\cite{yang1989eta,mark2020eta,moudgalya2020eta}. Although these states are separated into distinct symmetry sectors in the original Hubbard model, related extended Hubbard models realise them as exact quantum many-body scars. Together, these examples suggest that models solvable by the coordinate Bethe ansatz provide promising settings in which to search for exact or algebraically protected scar structures, and that combining orbit finding with the Bethe-ansatz identification of supporting eigenstates offers a natural systematic approach to this problem.
    
    Beyond suggesting new theoretical directions, the present results also point towards a possible experimental realisation: the scarred orbit identified here arises in a model that can be implemented on existing quantum-simulation platforms. Specifically, XXZ spin dynamics have already been realised in ultracold-atom experiments~\cite{jepsen2022long}, while the transverse field could be implemented through a global Rabi drive, making the non-integrable family studied here a promising target for experimental investigation. The initial state that seeds the scarred dynamics is a $\chi=2$ MPS and therefore admits a sequential state-preparation protocol of the type developed for matrix-product states in Ref.~\cite{schon2005sequential}.
    
    \section*{Acknowledgements}
    We acknowledge useful discussions with G\"okhan Yaln{\i}z, Aleksandr Zhabin, and Paul Fendley.
    M.L. acknowledges support by the Deutsche Forschungsgemeinschaft (DFG, German Research Foundation) under Germany’s Excellence Strategy – EXC-2111 – 390814868. 
    E.P. and M.S. acknowledge support by the Austrian Science Fund (FWF) \doi{10.55776/COE1} and the European Union–NextGenerationEU. We acknowledge support by the Erwin Schr\"o\-dinger International Institute for Mathematics and Physics (ESI). This research was supported in part by grant NSF PHY-2309135 to the Kavli Institute for Theoretical Physics (KITP).
    
    \section*{Appendix}
    \appendix
    
    \section{The extended tangent-space gradient-descent algorithm} 
    \label{app:algorithm} 
    
    The tangent-space gradient-descent algorithm used in this paper was introduced in Ref.~\cite{petrova2025finding}, where it is described in detail. Here, we discuss only the implementation of the energy-conservation constraint during the optimisation.
    
    Taking $|\psi\rangle$ to be normalised and choosing the tangent-space gauge such that $\langle\psi|\delta\psi\rangle=0$, we expand the energy expectation value to first order in the gradient-descent step:
    \begin{equation}
        \langle\psi+\delta\psi|\hat{H}|\psi+\delta\psi\rangle=E+2\,\mathrm{Re}\,\langle\delta\psi|\hat{H}|\psi\rangle+\mathcal{O}(\|\delta\psi\|^2),
        \label{eq:energy-expansion}
    \end{equation}
    where $E=\langle\psi|\hat{H}|\psi\rangle$. We require the linear term to vanish,
    \begin{equation}
        \mathrm{Re}\,\langle\delta\psi|\hat{H}|\psi\rangle=0.
        \label{eq:energy-constraint}
    \end{equation}
    Working in the MPS mixed-canonical form, we parametrise the tangent space at $|\psi\rangle$ by a basis $\{|V_i\rangle\}_{i=1}^{2(d-1)\chi^2}$ of linearly independent variations, where $\chi$ is the MPS bond dimension, $d$ is the local Hilbert-space dimension, and the factor of $2$ accounts for the real and imaginary components. The gradient-descent update is expanded in this basis,
    \begin{equation}
        |\delta\psi\rangle=\sum_{i=1}^{2(d-1)\chi^2}g^{(i)}|V_i\rangle,
    \end{equation}
    where the real coefficients $g^{(i)}$ are the components of the gradient vector $\vec{g}=(g^{(1)},\ldots,g^{(2(d-1)\chi^2)})$ produced by the algorithm of Ref.~\cite{petrova2025finding}. We compute the Hamiltonian contractions in the same basis,
    \begin{equation}
        \Delta E_i=\langle\psi|\hat{H}|V_i\rangle,
        \label{eq:dE-i}
    \end{equation}
    which enter the linear energy constraint through their real parts and form the row vector $\Delta\vec{E}=(\Delta E_1,\ldots,\Delta E_{2(d-1)\chi^2})$. The diagrammatic expression for $\Delta\vec{E}$ is
    \begin{equation}
        \adjincludegraphics[width=0.9\linewidth,valign=c,raise=0mm]{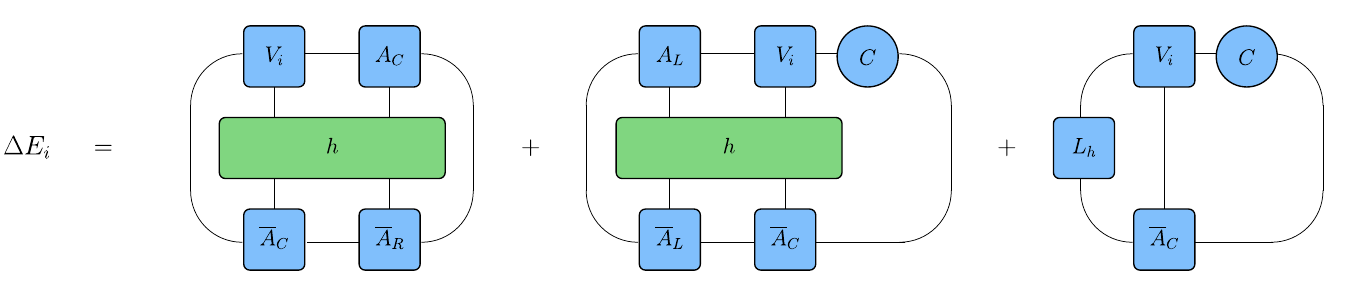}
        \label{eq:A-def}
    \end{equation}
    
    In this notation, the constraint~\eqref{eq:energy-constraint} becomes the single real linear equation $\Delta\vec{E}\cdot\vec{g}=0$, which is satisfied precisely when $\vec{g}$ lies in the $(2(d-1)\chi^2-1)$-dimensional null space of $\Delta\vec{E}$. Let $\{n_a\}_{a=1}^{2(d-1)\chi^2-1}$ denote an orthonormal basis of this null space in the space of coefficient vectors $\vec g$, with $n_a=(n_a^{(1)},\dots,n_a^{(2(d-1)\chi^2)})$. We define the reduced tangent basis as
    \begin{equation}
        |\tilde V_a\rangle \;=\; \sum_{i=1}^{2(d-1)\chi^2} n_a^{\,(i)}\,|V_i\rangle,
        \qquad a = 1,\dots, 2(d-1)\chi^2 - 1,
    \end{equation}
    so that every variation expressed in $\{|\tilde V_a\rangle\}$ preserves the energy to linear order by construction. The gradient-descent algorithm of Ref.~\cite{petrova2025finding} is then applied in this reduced tangent basis.
    
\begin{figure}
        \centering
        \includegraphics[width=0.7\linewidth]{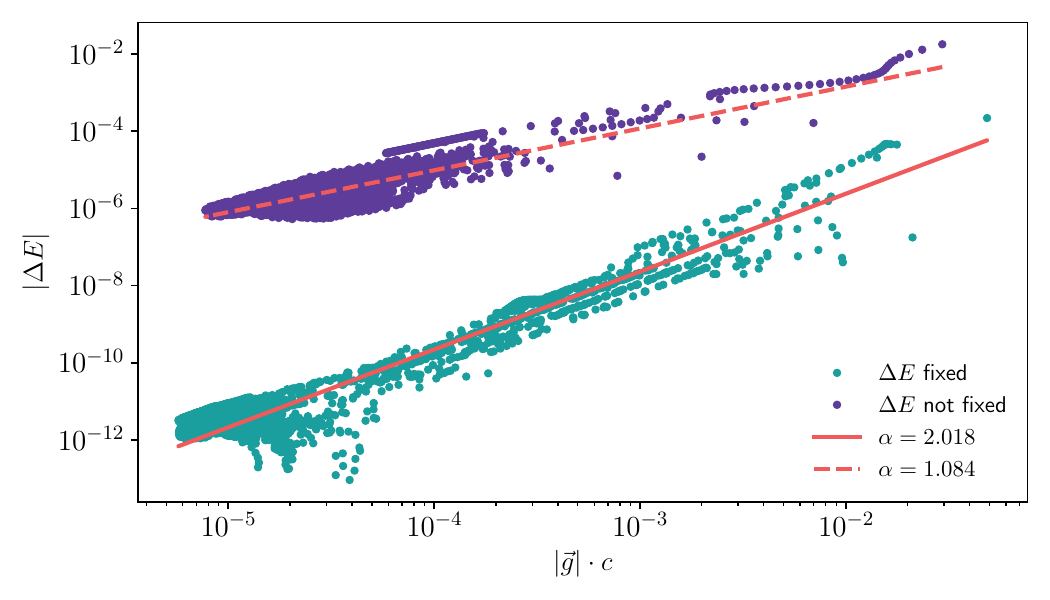}
        \caption{Energy change $|\Delta E|$ per gradient-descent step versus the update size $\|\vec g\|\,c$, where $c$ is a learning rate. Without projection (purple), the linear term in Eq.~\eqref{eq:energy-expansion} gives rise to a linear scaling of $|\Delta E|$ with the update size, as seen from the numerical exponent $\alpha\approx1.08$, defined by $|\Delta E|\propto(\|\vec g\|\,c)^\alpha$. After projection onto the null space of $\Delta\vec{E}$ (teal), the linear contribution vanishes and only the quadratic remainder survives, giving $\alpha\approx2.02$.}
        \label{fig:dE_test}
    \end{figure}

    We verify numerically that the projection removes the linear term in Eq.~\eqref{eq:energy-expansion}. Figure~\ref{fig:dE_test} shows the energy change $|\Delta E|$ per gradient-descent step as a function of the update size $\|\vec g\|\,c$, where $c$ is a learning rate. Without the projection (purple), $|\Delta E|$ scales linearly with the update size, with a power-law scaling exponent $\alpha\approx 1$, reflecting the first-order term $2\,\mathrm{Re}\,\langle\delta\psi|\hat H|\psi\rangle$. After projecting $\vec g$ onto the null space of $\Delta\vec E$ (teal), this term vanishes by construction, and only the $\mathcal{O}(\|\delta\psi\|^2)$ remainder survives, giving a power-law scaling exponent $\alpha\approx 2$. The constraint is therefore satisfied to numerical precision at every iteration.
    
    To generate a family of orbits at different energies, we proceed iteratively. Starting from a converged periodic orbit at energy $E$, we apply a small perturbation chosen to move the trajectory towards a neighbouring energy shell at $E+\delta E$ and use the perturbed trajectory as the initial condition for a new run of the energy-conserving algorithm. The optimisation converges to a new periodic orbit at energy $E+\delta E$, which then serves as the seed for the next iteration. Repeating this procedure yields a family of orbits spanning a wide range of energies.
    
    \section{Solution of the Bethe equations} 
    \label{app:Bethe}
    
    We solve Eq.~\eqref{eq:loBAE} for the complex momenta $\{k_j\}$ by Newton iteration on the residual $\vec{F}(\vec{k})=\vec{k}-(2\pi/L)\vec{I}-(1/L)\vec{\theta}(\vec{k})$, where $\theta_j(\vec{k})=\sum_{l\neq j}\theta(k_j,k_l)$. The Jacobian $\partial F_i/\partial k_j$ is evaluated numerically using finite differences in the complex variables $k_j$, and the update is damped, $\vec{k}\to\vec{k}-\eta\mathbf{J}^{-1}\vec{F}$, with $\eta=0.05$, until $\|\vec{F}\|_\infty<10^{-9}$.
    
    The initial guess and the structure of the solution depend on the quantum-number pattern:
    \begin{itemize}
    \item \textit{Real roots.} 
    A quantum number $I_j$ that appears only once in $\{I_l\}$ corresponds to a real magnon momentum $k_j\in\mathbb{R}$. For each such index, we initialise $k_j=2\pi I_j/L$.
    \item \textit{Two-string at $I=L/2$.} 
    A repeated quantum number $I_j=I_{j+1}=L/2$ signals a two-string (bound magnon pair). In the ferromagnetic regime, the analytic two-string ansatz is $k_\pm=\pi\pm i\eta$ with $\cosh\eta=|\Delta/J|$. Numerically, we initialise the pair as $k_\pm=\pi\pm ir$, where $r$ is a small random imaginary perturbation, and let the iteration converge to the corresponding value of $\eta$. When the same quantum number $L/2$ appears $2n$ times, the iteration converges to a stacked-string configuration with $n$ symmetric pairs of imaginary parts $\pm r_1,\pm r_2,\ldots,\pm r_n$, all sharing the real part $\pi$.
    \end{itemize}
    
    For a given solution $\{k_j\}$ of the Bethe equations, we construct the coordinate Bethe wave function in the sector $x_1<\cdots<x_M$ using the symmetric convention~\cite{karbach1998introduction1,franchini2017introduction},
    \begin{equation}
        \psi(x_1,\ldots,x_M)
        \;=\;
        \sum_{\sigma\in\mathcal{S}_M}
        \exp\!\Bigl[
            i\sum_l k_{\sigma(l)}\,x_l
            +
            \frac{i}{2}
            \sum_{j<l}
            \theta\!\bigl(k_{\sigma(j)},k_{\sigma(l)}\bigr)
        \Bigr].
    \end{equation}
    where the XXZ scattering phase is defined through~\cite{karbach1998introduction1}
    \begin{equation}
        e^{i\theta(k_1,k_2)}
        =
        -\,\frac{e^{i(k_1+k_2)} + 1 - 2\Delta J^{-1}\,e^{ik_1}}
                  {e^{i(k_1+k_2)} + 1 - 2\Delta J^{-1}\,e^{ik_2}}.
        \label{eq:S-matrix}
    \end{equation}
    
    \section{XYZ scars}
    \label{app:xyz}
    
    \begin{figure}
        \centering
        \includegraphics[width=1.\linewidth]{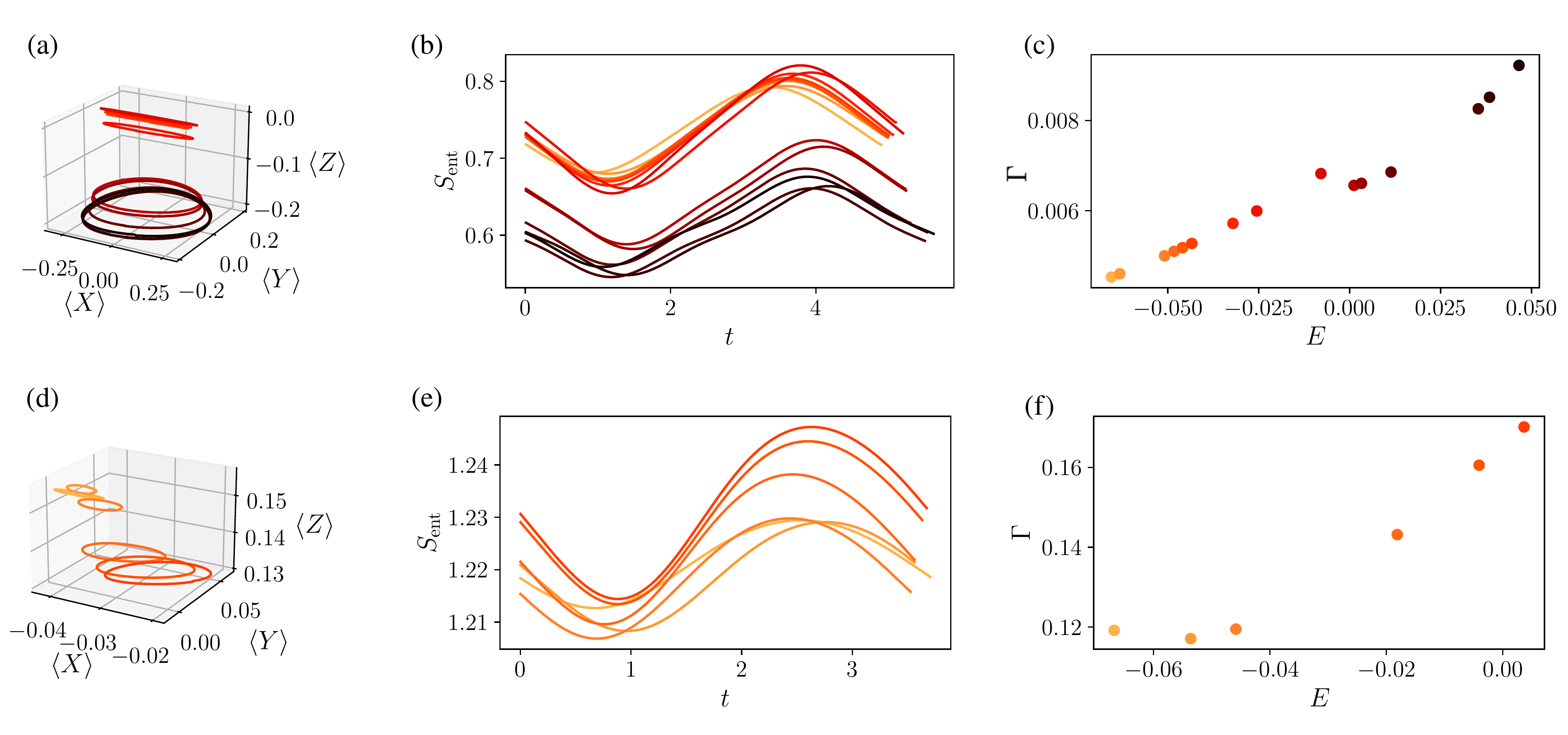}
        \caption{Energy-tracked families of $\chi = 3$ (top, a--c) and $\chi = 4$ (bottom, d--f) periodic orbits in the XYZ chain at $J_x = 1$, $J_y = 0.8$, $\Delta = -0.5$. Colour encodes energy, from low (light) to high (dark), along each family. (a, d) Orbits as seen in expectation values of local observables over one period. (b, e) Entanglement entropy $S_{\rm ent}(t)$ along each orbit. (c, f) Integrated leakage $\Gamma$ as a function of orbit energy.}
        \label{fig:app3}
    \end{figure}

    \begin{figure}
        \centering
        \includegraphics[width=1.\linewidth]{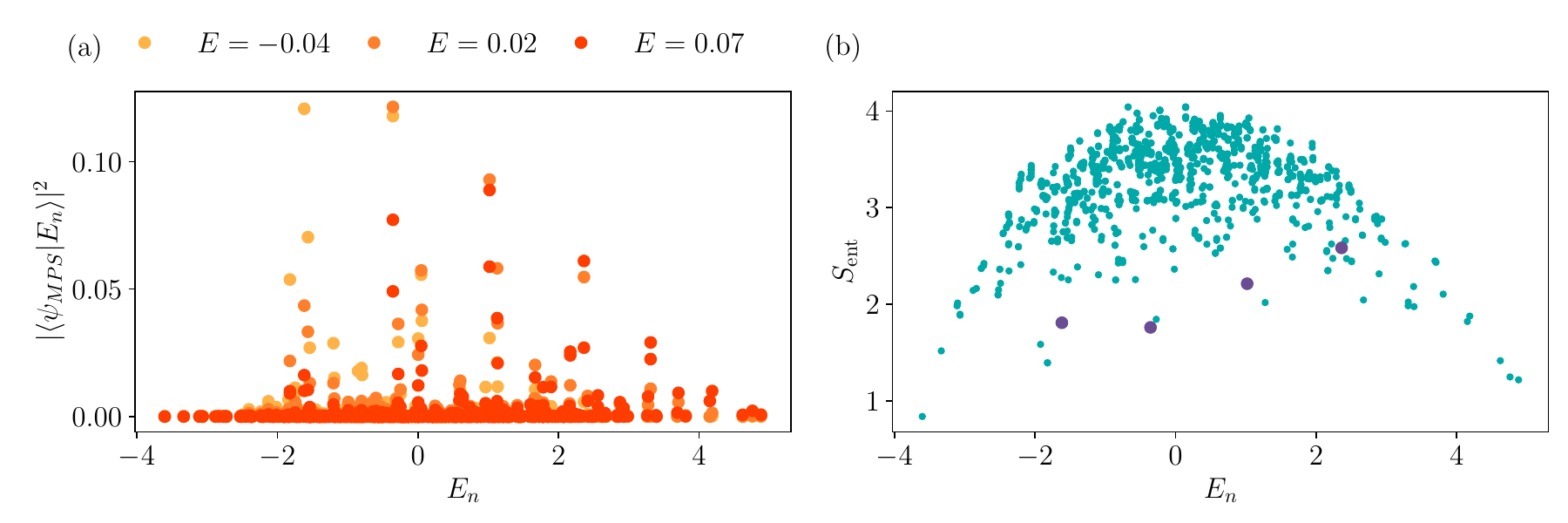}
        \caption{(a) Overlap of the orbit with the eigenstates of the XYZ Hamiltonian for $L = 14$ in the zero momentum sector. The energy densities shown here are computed from the finite-size eigenstates at $L=14$, which is why they differ from the energy densities shown in Fig.~\ref{fig:app3}. (b) Entanglement entropy of the eigenstates. The violet dots correspond to the eigenstates contributing most strongly to the orbit in panel (a).}
        \label{fig:app3_2}
    \end{figure}

    This Appendix presents additional data for the $\chi=3$ and $\chi=4$ orbits in the XYZ model. Figure~\ref{fig:app3}(a, d) shows the behaviour of local observables for the $\chi=3$ and $\chi=4$ orbits, respectively, tracked across the same range of energies as in Sec.~\ref{sec:scars}. The corresponding entanglement entropy and integrated leakage $\Gamma$ along each orbit are reported in panels (b,c) for $\chi=3$ and in (e,f) for $\chi=4$. The $\chi=3$ family inherits the same counter-intuitive trend observed at $\chi=2$: higher-energy orbits have lower entanglement entropy, while the integrated leakage grows with energy.

    Figure~\ref{fig:app3_2}(a,b) traces this behaviour back to the structure of the supporting eigenstates. As one moves along the family towards higher energies, the orbit's weight shifts towards higher-energy eigenstates and greater entanglement. The variational manifold remains pinned at $\chi=3$, so the additional entanglement cannot be fully represented within the MPS manifold; this reduces the entanglement visible in the MPS while increasing the projection error, producing the simultaneous decrease in $S_{\rm ent}$ and increase in $\Gamma$ observed in Fig.~\ref{fig:app3}.

    \section{XXZ scars} 
    \label{app:xxz}

    \begin{figure}
        \centering
        \includegraphics[width=1.\linewidth]{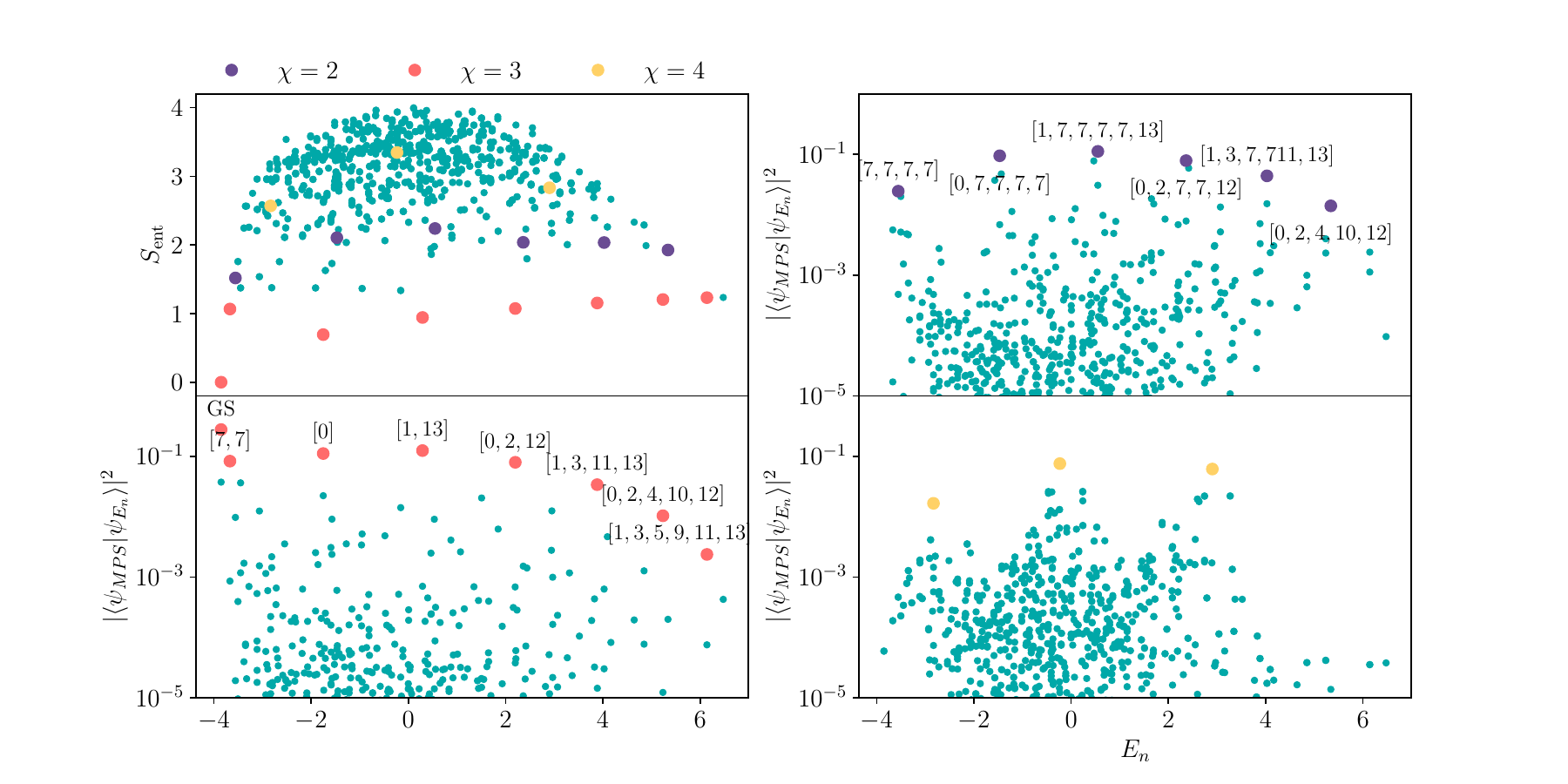}
        \caption{Bethe-quantum-number identification of the scars at $\chi=2,3,4$ for the XXZ chain at $L=14$ and $\Delta=-1.1$. Top-left panel: entanglement entropy $S_{\mathrm{ent}}$ of all zero-momentum eigenstates, with the dominant supporting eigenstates of the $\chi=2$ (violet), $\chi=3$ (peach), and $\chi=4$ (yellow) trajectories highlighted. The remaining three panels show the overlap $|\langle\psi_{\mathrm{MPS}}|\psi_{E_n}\rangle|^2$ of each MPS orbit ($\chi=2,3,4$ in the top-right, bottom-left, and bottom-right panels, respectively) with the eigenstates of the Hamiltonian, with the Bethe quantum numbers $\{I_j\}$ of the dominant eigenstates annotated. The $\chi=3$ ladder reproduces the $\chi=2$ dressing structure, but without the $I=L/2$ core.}
        \label{fig:all_scars}
    \end{figure}
    
\begin{figure}[t!]
        \centering
        \includegraphics[width=1.\linewidth]{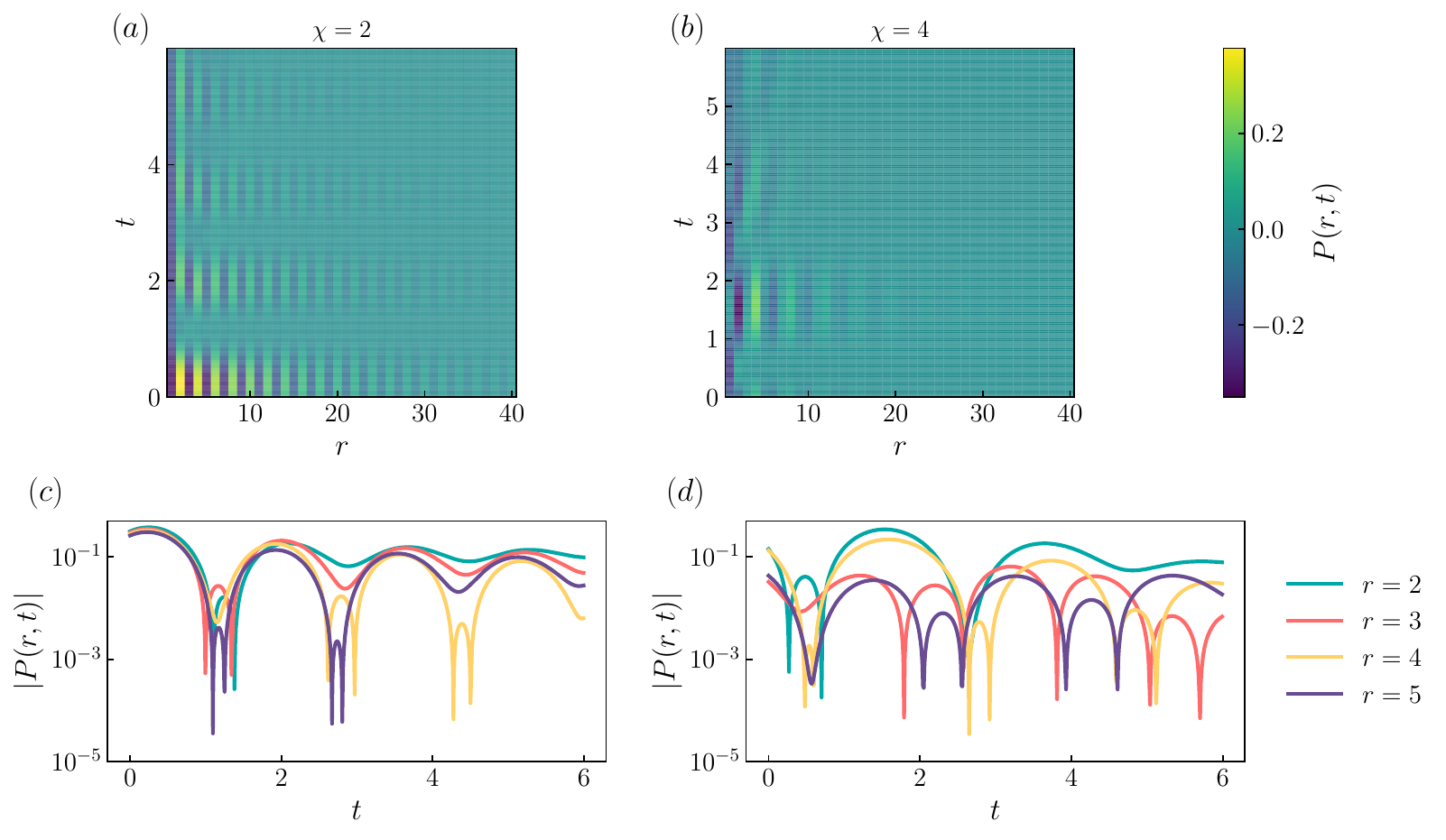}
        \caption{(a,b) Dynamics of the observable $P(r,t)$ in TEBD simulations for the $\chi=2$ and $\chi=4$ orbits, respectively. (c,d) Time traces extracted from panels (a,b). In panel (c), the dynamics is similar for all values of $r$, differing primarily by an overall sign. In panel (d), the $r=2$ and $r=4$ traces show similar behaviour, whereas the $r=3$ and $r=5$ traces have a smaller average amplitude and a qualitatively different time dependence.}
        \label{fig:chi4}
    \end{figure}

    As discussed in the main text, at system size $L=14$, the orbits with $\chi=2,3$ found by the algorithm are supported by the same ladder of scarred eigenstates, with the only difference being the size of the $I=L/2$ string core. Figure~\ref{fig:all_scars} extends Fig.~\ref{fig:bethe-id} of the main text by showing the entanglement entropy of eigenstates together with the full overlap structure for each bond dimension; the $\chi=3$ ladder is similar to the $\chi=2$ ladder, but with the bound-state core removed. The $\chi=4$ orbit is more delicate. The solver for the Bethe equations is sensitive to the initial guess for the momenta. For $\chi=4$, we have not been able to identify a quantum-number assignment that converges to a Bethe state with significant overlap with the orbit, and we therefore report only the Bethe-side analysis of $\chi=2,3$ in the main text. The only indication of a distinct microscopic structure comes from the behaviour of the correlator $P(r,t)$ in Fig.~\ref{fig:chi4}. The repeating pattern in the $r$ direction suggests a different organisation of the Bethe quantum numbers than in the $\chi=2$ orbit.

    Guided by the ladder and tower structures identified at smaller $L$ in Sec.~\ref{sec:bethe}, we run the Bethe solver on system sizes between 8 and 30 and restrict the search to quantum-number patterns containing at least one such core pair. We construct the corresponding eigenstates directly from the Bethe ansatz and evaluate their overlap with the MPS scar orbit. Searching over all patterns of this form, we identify, for each system size, the eigenstate of largest overlap. The resulting quantum numbers are listed in Table~\ref{tab:bethe-max-overlap}, and the overlaps themselves are plotted in Fig.~\ref{fig:bethe-id}(b) of the main text.
    
    \begin{table}
    \centering
    \begin{tabular}{c|l}
    \hline\hline
    $L$  & $\{I_j\}$ \\
    \hline
    8  & $[0,\,4\!\times\!2]$ \\
    10 & $[1,\,5\!\times\!2,\,9]$ \\
    12 & $[1,\,6\!\times\!2,\,11]$ \\
    14 & $[1,\,7\!\times\!4,\,13]$ \\
    16 & $[1,\,8\!\times\!4,\,15]$ \\
    18 & $[0,\,2,\,9\!\times\!4,\,16]$ \\
    20 & $[0,\,2,\,10\!\times\!6,\,18]$ \\
    22 & $[0,\,2,\,11\!\times\!6,\,20]$ \\
    24 & $[1,\,3,\,12\!\times\!6,\,21,\,23]$ \\
    26 & $[0,\,2,\,12,\,13\!\times\!6,\,14,\,24]$ \\
    28 & $[1,\,3,\,13,\,14\!\times\!6,\,15,\,25,\,27]$ \\
    30 & $[1,\,3,\,14,\,15\!\times\!6,\,16,\,27,\,29]$ \\
    \hline\hline
    \end{tabular}
    \caption{Bethe quantum numbers $\{I_j\}$ for the eigenstates with maximum overlap with the scar orbit, for system sizes $L=8,10,\ldots,30$ (XXZ chain at $\Delta=-1.1$, $\chi=2$ orbit). The notation $a\!\times\!n$ denotes $n$ occurrences of the Bethe quantum number $I=a$. Each row decomposes into a core consisting of $2n$ copies of $I=L/2$ together with zero-momentum dressing units.}
    \label{tab:bethe-max-overlap}
    \end{table}
    
\begin{figure}
        \centering
        \includegraphics[width=1\linewidth]{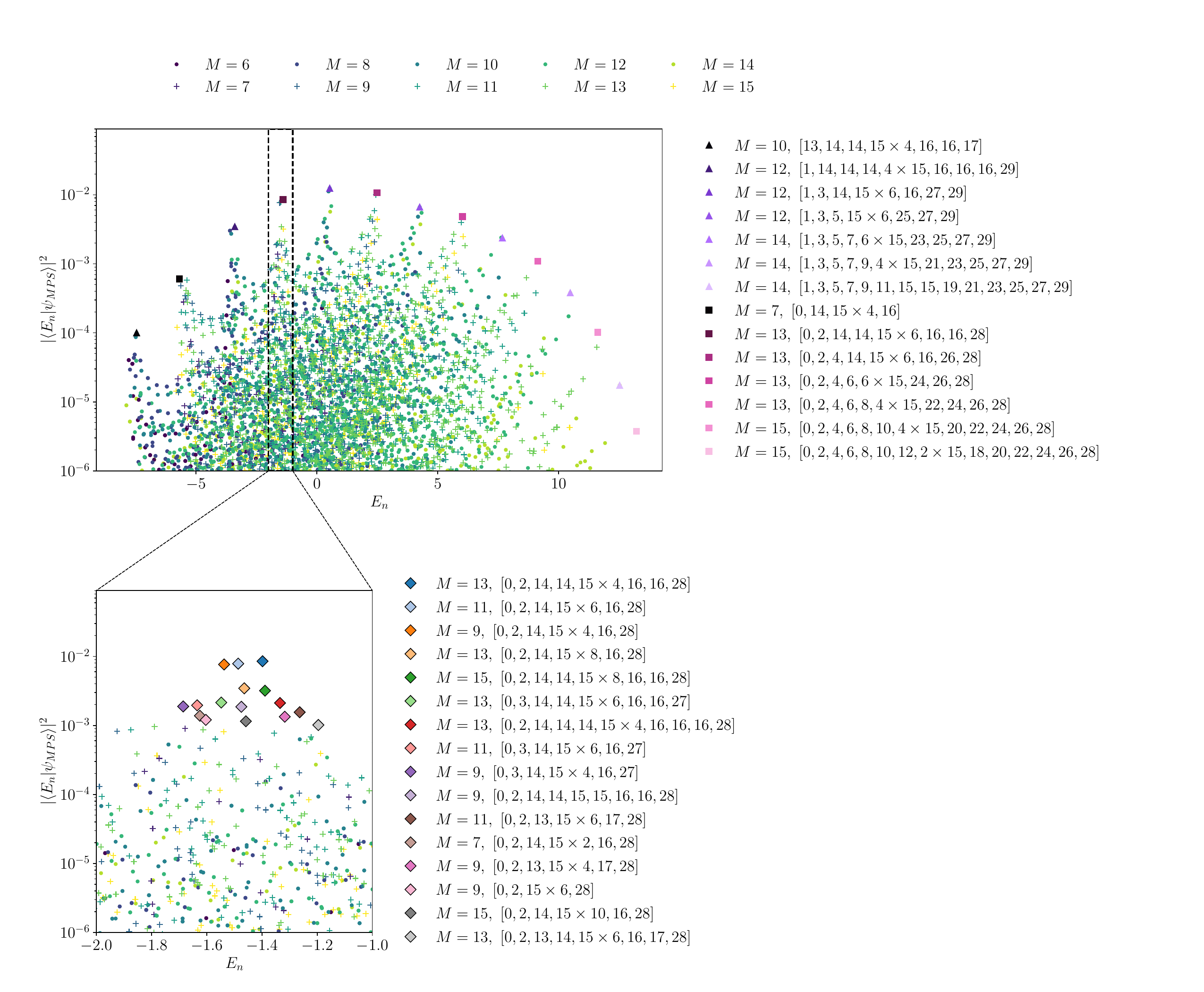}
        \caption{Overlap $|\langle\psi_\mathrm{MPS}|\psi_{E_n}\rangle|^2$ of the $\chi=2$ scar orbit with the eigenstates of the XXZ chain at $L=30$, $\Delta=-1.1$, grouped by magnetisation sector $M$ (different colours, see top legend) and labelled by the Bethe quantum numbers $\{I_j\}$ of the dominant contributions (colour and shape, see right legend). The tower structure identified at $L=14$ in Fig.~\ref{fig:all_scars} persists: the eigenstates organise into towers of nearly degenerate states (same dressing, different bound-state core size). Every dominant eigenstate can be decomposed into an $I=L/2$ core together with a zero-momentum dressing of symmetric pairs $\{a,L-a\}$ (and, in odd-$M$ sectors, an additional $I=0$ root).}
        \label{fig:scar30}
    \end{figure}

    For system size $L=30$, we report the overlap of the scar orbit with the full set of solved Bethe states (6515 of 108528) in Fig.~\ref{fig:scar30}, where the tower-and-ladder structure is clearly visible: the eigenstates organise into towers of nearly degenerate states (same dressing, different bound-state core size), and the towers themselves form an approximately equidistant ladder. The state with the largest overlap inside each tower is marked. Adding a single dressing unit shifts the energy of the supporting state by approximately the same amount, in agreement with the equidistant-ladder picture. However, once several dressing units are present, the inter-rung spacing begins to contract, and the overlap amplitude decreases, reflecting the residual scattering between dressing magnons that would otherwise be suppressed by $1/L$ in a larger system.
    
    \section{TEBD simulations}
    \label{app:tebd}

    \begin{figure}[b!]
    	\centering
    	\includegraphics[width = 0.5\linewidth]{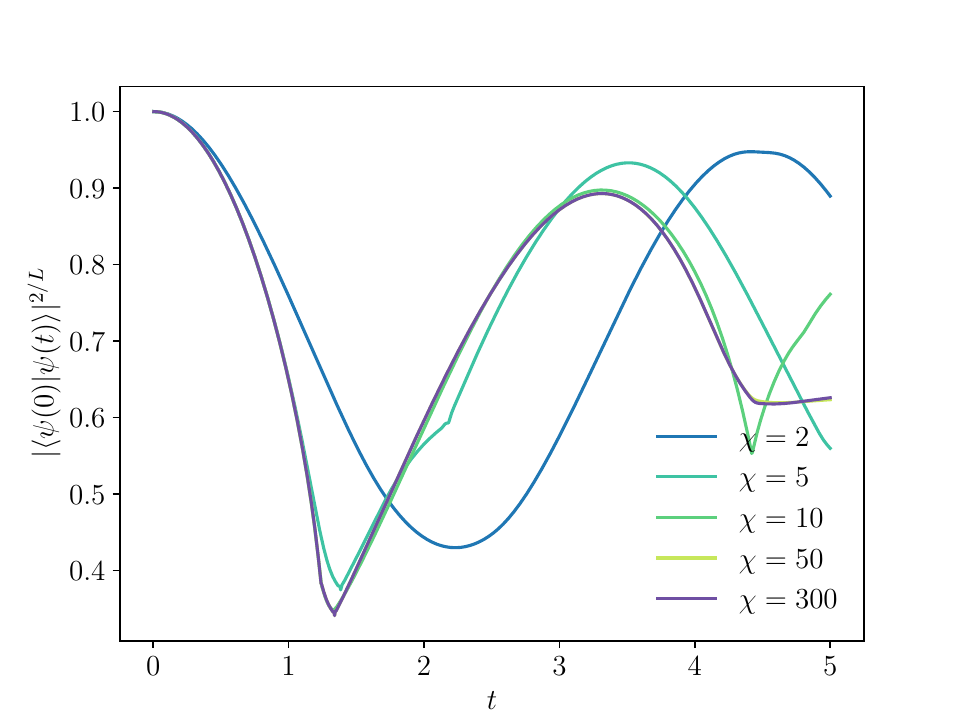}
    	\caption{Convergence of the revival period with the TEBD bond dimension for $L=200$, $\Delta=-1.1$, starting from the $\chi=2$ scar state. We plot the revival fidelity $|\langle\psi(0)|\psi(t)\rangle|^{2/L}$ for $\chi_{\mathrm{TEBD}}=2,5,10,50,300$. The position of the first revival drifts towards shorter times as $\chi_{\mathrm{TEBD}}$ increases and saturates above $\chi_{\mathrm{TEBD}}\!\sim\!50$, indicating an apparent period mismatch between the TEBD dynamics and the $\chi=2$ TDVP orbit.}
    	\label{fig:App_period}
    \end{figure}
    
    The results shown in Figs.~\ref{fig:int_b} and~\ref{fig:revival1} are obtained using the TEBD algorithm implemented in the \textsc{ITensor} library~\cite{ITensor}. We use a chain of length $L=200$ with open boundary conditions and a fixed time step $\delta t=0.005$ throughout. After each two-site gate, the bond is truncated by singular-value decomposition, retaining at most $\chi_{\max}=300$ singular values and imposing a discarded-weight cut-off of $\varepsilon_{\rm cut}=10^{-20}$. The initial state is the $\chi=2$ MPS produced by the orbit-finding algorithm of Sec.~\ref{sec:scars}, embedded into a chain of length $L=200$. The bond dimension is allowed to grow freely up to $\chi_{\max}$ as entanglement is generated during the evolution. All singular values beyond this threshold are truncated.
    
    Fig.~\ref{fig:App_period} tracks the dependence of the revival period of the $\chi=2$ orbit on $\chi_{\max}$. The revival period drifts visibly at small bond dimensions and saturates as $\chi_{\max}$ is increased; the residual shift at $\chi_{\max}=100$ provides our estimate of the bond-dimension uncertainty in the period quoted in Sec.~\ref{sec:int_b}.
    
    \bibliography{bib}
    
\end{document}